\documentclass[final,3p,times]{elsarticle}

\usepackage[T1]{fontenc}
\usepackage[utf8]{inputenc}
\usepackage{amsmath,amssymb,amsfonts}
\usepackage{bm}
\usepackage{booktabs}
\usepackage{longtable}
\usepackage{array}
\usepackage{siunitx}
\usepackage{graphicx}
\usepackage{placeins}
\usepackage{xcolor}
\usepackage{listings}
\usepackage{hyperref}
\hypersetup{hidelinks}
\journal{Computer Physics Communications}

\lstdefinelanguage{Julia}{
  morekeywords={function,struct,mutable,if,else,elseif,for,while,begin,end,return,using,include,true,false,nothing,try,catch,const,import},
  sensitive=true,
  morecomment=[l]{\#},
  morestring=[b]"
}

\newcommand{\code}[1]{\texttt{\detokenize{#1}}}
\newcommand{\kvec}{\mathbf{k}}
\newcommand{\qvec}{\mathbf{q}}
\newcommand{\Rvec}{\mathbf{R}}
\newcommand{\dk}{\mathrm{d}\mathbf{k}}
\newcommand{\ii}{\mathrm{i}}
\newcommand{\Tr}{\operatorname{Tr}}

\begin{document}

\begin{frontmatter}

\title{WannierNLQG: A Julia package for nonlinear optical responses and quantum geometry from Wannier tight-binding models}

\author[label1]{Zhuocheng Lu}
\author[label1,label2]{Zhichao Guo}
\author[label1]{Yuanyuan Xu}
\author[label1]{Jiacheng Yao}
\author[label1]{Hua Wang\corref{cor1}}
\ead{daodaohw@zju.edu.cn}
\cortext[cor1]{Corresponding author.}
\address[label1]{Center for Quantum Matter, School of Physics, Zhejiang University, Hangzhou 310058, China}
\address[label2]{Department of Physics, City University of Hong Kong, Kowloon 999077, Hong Kong, China}

\begin{abstract}
Nonlinear optical responses and quantum geometry are central to understanding electronic phenomena in quantum materials and are deeply intertwined. Building on this dual focus, we introduce WannierNLQG, an extensible Julia framework that enables the computation of nonlinear optical responses and quantities in quantum geometry directly from Wannier tight-binding models. Its gauge-consistent, degeneracy-aware architecture provides a common foundation for extending calculations of nonlinear response across arbitrary perturbative orders and for incorporating additional quantities in quantum geometry.
The current release evaluates ordinary, spin, and photon drag injection and shift currents, both as Brillouin-zone integrals and on k slices. For shift current, a unified interface exposes four complementary formulations: the conventional method, projector trace, generalized Wilson loop, and a geometric loop defined at finite momentum, with the last three explicitly accommodating degenerate subspaces. The framework also provides k-resolved Berry curvature, quantum metric, and their multipoles, together with additional quantities in quantum geometry and quantities coupling momentum and spin.
Complementing these core capabilities, experimental symmetry workflows support Wannier construction adapted to symmetry, symmetrization of tight-binding models and real-space operators, and reduction of spatially uniform response integrals to irreducible k-point orbits and invariant tensor components.
The documentation details formula conventions, the TaskConfig interface, and auditable output formats, and includes GeS and bilayer MoS$_2$ case studies illustrating method comparison, degeneracy handling, analysis of quantum geometry, and photon drag responses.
Together, these capabilities support first-principles calculations of nonlinear response and quantum geometry in realistic multiband materials, connecting quantum geometry with quantitative materials modeling for a broad range of optoelectronic, spintronic, and photovoltaic applications.
\end{abstract}

\begin{keyword}
Wannier interpolation \sep nonlinear optics \sep shift current \sep injection current \sep quantum geometry \sep photon drag \sep symmetry-adapted Wannier functions
\end{keyword}

\end{frontmatter}

\section{Program summary}

\begin{longtable}{@{}p{0.30\linewidth}p{0.64\linewidth}@{}}
\toprule
Program title & WannierNLQG \\
CPC Library link to program files & To be added by the CPC Library after acceptance \\
Developer's repository & \url{https://github.com/ZhuochengLu/WannierNLQG} \\
Licensing provisions & GNU General Public License version 2 only (GPL-2.0-only) \\
Programming language & Julia \\
External routines/libraries & FFTW.jl, MPI.jl, HDF5.jl, JSON3.jl, EzXML.jl, and Spglib.jl. \\
Supplementary material & A Mendeley Data reproducibility archive, Version 1, is publicly available at \url{https://doi.org/10.17632/v2s2j336vh.1}. \\
Nature of problem & Nonlinear optical response calculations require dense Brillouin-zone sampling and gauge-consistent treatment of degenerate band subspaces. Existing packages do not systematically support the broader family of quantities in quantum geometry, including multipoles, higher-rank tensors, and objects coupling momentum and spin. Software support is also lacking for responses at finite momentum based on Wannier models, where optical matrix elements connect electronic states at distinct crystal momenta. \\
Solution method & Projector, Wilson loop, and geometric loop formulations provide gauge-consistent subspace treatments of band degeneracies. Within the same Wannier framework, geometric quantities are evaluated systematically, including multipole moments, higher-order tensors, and quantities coupling momentum and spin. Furthermore, optical matrix elements connecting distinct crystal momenta enable nonlinear optical response calculations at finite momentum, including photon drag shift and injection currents. \\
\bottomrule
\end{longtable}

\section{Introduction}

Maximally localized Wannier functions and disentanglement algorithms provide a compact route from first-principles electronic structures to high-accuracy interpolation of realistic multiband Hamiltonians and matrix elements over dense Brillouin-zone meshes \cite{Marzari1997MLWF,Souza2001EntangledWannier}. The Wannier90 code implements these procedures and generates associated tight-binding representations. Its community releases established these methods as standard computational infrastructure \cite{Mostofi2008Wannier90,Mostofi2014Wannier90Update,Pizzi2020Wannier90}. Wannier interpolation is now central to topology, quantum geometry, transport, and optical response calculations \cite{KingSmith1993Polarization,Resta1994Polarization,Wang2006AHCWannier,Yates2007WannierInterpolation}. Around this foundation, standardized interfaces and data formats connect first-principles codes, Wannierization engines, and downstream tools into an interoperable ecosystem encompassing topological analysis, electron--phonon calculations, beyond-DFT methods, and increasingly automated and high-throughput workflows \cite{Gresch2017Z2Pack,Lee2023EPW,Vitale2020AutomatedWannierisation,Qiao2023ProjectabilityDisentanglement,Marrazzo2024WannierEcosystem}. For example, WannierTools supports topological-material analysis \cite{Wu2018WannierTools}, WannSymm supports symmetry analysis and symmetrization of Wannier models \cite{Zhi2022WannSymm}, and WannierBerri supports Wannierization and symmetry-aware workflows alongside efficient interpolation and Brillouin-zone integration of transport and optical response quantities \cite{Tsirkin2021WannierBerri,WannierBerriDocumentation}.

Despite the maturity of this ecosystem, three gaps persist. First, a central numerical difficulty remains in nonlinear optical response calculations. The shift current provides a paradigmatic example. Although the shift vector itself is gauge invariant, its standard expression combines diagonal Berry connections with the momentum derivative of the phase of a gauge-dependent interband optical matrix element, so direct finite differences of these intermediate quantities are not gauge stable. Conventional Wannier interpolation schemes circumvent this instability by replacing the direct derivative with a sum-over-states formulation within the Wannier subspace, thereby also avoiding the band-truncation errors of direct Bloch-band sums \cite{Wang2017NLOWannier,IbanezAzpiroz2018ShiftWannier}. This advance, however, does not remove the underlying single-band limitation: at band degeneracies, eigenvectors within the degenerate manifold are defined only up to arbitrary unitary rotations, rendering individual matrix elements and their derivatives basis dependent. Regularization removes divergences from vanishing energy denominators but does not recover the collective contribution of the degenerate subspace. By contrast, projector, Wilson loop, and geometric loop formulations offer complementary routes built directly from gauge-invariant projectors or closed-loop objects: the momentum derivatives of these objects can be evaluated by finite differences, and their isolated-band expressions generalize naturally to degenerate subspaces \cite{Wang2022WilsonLoop,Avdoshkin2025MultistateGeometry,Guo2025Projector,Lu2025GiantPhotonDrag}.

Second, and equally important, no existing package systematically computes the broader family of quantities in quantum geometry that have become central not only to modern theory of nonlinear response but also to a rapidly expanding set of condensed-matter phenomena, including superconductivity and strongly correlated states \cite{torma2022superconductivity,liu2025quantum}. Berry curvature and the quantum metric appear in isolation across several codes, but their multipole moments, higher-rank geometric tensors, and generalizations coupling momentum and spin (objects increasingly central to nonlinear Hall and orbital-magnetic phenomena) are typically implemented case by case or not at all. A unified, gauge-consistent, and degeneracy-aware framework that both supports the alternative shift current formulations above and exposes these geometric quantities on the same footing is therefore essential for gauge-consistent calculations of nonlinear response and quantum geometry in realistic multiband materials.

Third, the major software packages for calculating nonlinear optical responses from Wannier tight-binding models are currently restricted to the spatially uniform \(q=0\) limit \cite{Wang2017NLOWannier,Tsirkin2021WannierBerri,WannierBerriDocumentation}. Although the corresponding responses may involve interband matrix elements, the physical optical matrix elements entering them remain diagonal in crystal momentum, with the coupled Bloch states evaluated at the same \(\kvec\). Photon drag nonlinear optical responses are qualitatively different: finite photon momentum connects states at \(\kvec_\alpha=\kvec-\qvec/2\) and \(\kvec_\beta=\kvec+\qvec/2\), requiring optical matrix elements and objects in quantum geometry that connect distinct crystal momenta \cite{Shi2021PhotonDrag,Xie2025PhotonDragBPVE,Lu2025GiantPhotonDrag}. The construction and evaluation of these matrix elements at finite \(\qvec\), which are off-diagonal in \(k\), and their associated quantities in quantum geometry therefore provide the foundation for a general Wannier-based treatment of nonlinear optical responses at finite momentum.

\begin{figure*}[!tbp]
\centering
\includegraphics[width=\textwidth]{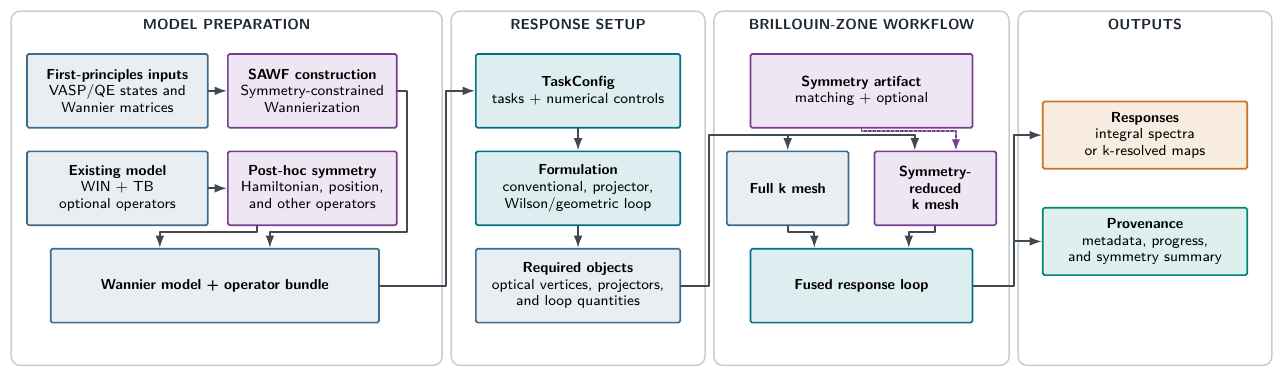}
\caption{Symmetry-aware computational workflow of WannierNLQG. Model Preparation produces a Wannier model and operator bundle through either representation-constrained SAWF construction from first-principles inputs or post-hoc symmetrization of an existing WIN and tight-binding model with optional operators. In Response Setup, \code{TaskConfig} selects the response formulation and the required optical vertices, projectors, or loop quantities. The Brillouin-zone stage uses the full k mesh by default. When a matching symmetry artifact is supplied, supported \(q=0\) Integral tasks may instead use the symmetry-reduced k mesh, implemented using representatives of irreducible k-point orbits, orbit multiplicities with full-mesh normalization, and invariant-tensor reconstruction. Both paths enter the fused response loop. The outputs comprise integral spectra or k-resolved maps together with metadata, progress records, and a symmetry summary; K-slice and finite-\(q\) tasks follow the full-mesh path.}
\label{fig:workflow}
\end{figure*}

The development of WannierNLQG draws on these advances and on a sustained body of prior work on nonlinear optical and photocurrent responses \cite{Guo2025Projector,Lu2025GiantPhotonDrag,Wang2017GiantSHG,Wang2019FerroicityPhotocurrent,Wang2019FerroelectricNAHE,Wang2020SwitchablePhotocurrent,Xu2021BulkSpinPhotovoltaic,Hu2025PhononQuantumGeometry,Wang2026GeodesicShiftVector}. Its gauge-consistent, degeneracy-aware architecture provides a common foundation for extending calculations of nonlinear response across arbitrary perturbative orders and for incorporating additional quantities in quantum geometry. As summarized in Fig.~\ref{fig:workflow}, a common Wannier representation connects model preparation, task-dependent geometric and optical objects, and response evaluation. This organization brings subspace treatments, broader analysis of quantum geometry, and matrix elements at finite momentum into a shared computational workflow.

The current release evaluates ordinary injection and shift currents, their spin current counterparts, and photon drag injection and shift currents, both as Brillouin-zone integrals and on k slices. For shift current, a common interface provides the conventional, projector trace, and generalized Wilson loop formulations, together with a geometric loop formulation at finite momentum; the latter three use gauge-covariant subspace treatments for degenerate bands. The framework also provides k-resolved Berry curvature, the quantum metric, and their multipole moments, together with higher-order quantum geometric tensors and quantities coupling momentum and spin. A common \code{TaskConfig}, output, and metadata workflow keeps response spectra, local diagnostics, formulation choices, and numerical provenance consistently aligned across tasks.

Symmetry-aware expert workflows enter at three distinct stages. Representation-constrained SAWF construction \cite{Sakuma2013SAWF} and post-hoc group averaging of existing real-space models and auxiliary operators act during model preparation. For supported spatially uniform response integrals, the symmetry-reduced k mesh in Fig.~\ref{fig:workflow} consists of representatives of irreducible k-point orbits. Their multiplicities retain full-mesh normalization and are combined with invariant-tensor reconstruction \cite{Togo2024Spglib,Shinohara2023MagneticSymmetry}. These workflows act on different mathematical objects and can be used independently or in sequence. Within the present implementation, they share a common provenance model while remaining operationally distinct.

The following sections develop and assess these capabilities. Section~\ref{sec:theory} establishes the Wannier representation and matrix element conventions and presents the implemented response formulations at zero and finite photon momentum. Section~\ref{sec:software} describes the architecture and public interface, supported calculations, model preparation and symmetry workflows, and reproducible outputs. Section~\ref{sec:examples} examines GeS and 2H-bilayer MoS$_2$ examples of method validation, degeneracy handling, analysis of quantum geometry, and responses at finite photon momentum, together with a frozen comparison of the full conventional shift current tensor across three materials at matched nominal compute unit counts. Section~\ref{sec:conclusions} summarizes the present capabilities and outlines future development.

\section{Theory and methods}
\label{sec:theory}

\subsection{Wannier representation and matrix elements}
\label{sec:wannier_tb}

The starting point is a Wannier90-style tight-binding representation. Following standard Wannier interpolation conventions \cite{Marzari1997MLWF,Souza2001EntangledWannier,Yates2007WannierInterpolation,Pizzi2020Wannier90}, the Wannier-gauge Bloch basis and Hamiltonian are
\begin{equation}
|u_{n\kvec}^{(\mathrm{W})}\rangle
=
\sum_{\Rvec}
e^{-\ii\kvec\cdot(\hat{\mathbf{r}}-\Rvec)}
|\Rvec n\rangle ,
\label{eq:wannier_gauge_basis}
\end{equation}
\begin{equation}
H_{nm}^{(\mathrm{W})}(\kvec)
=
\sum_{\Rvec}
e^{\ii\kvec\cdot\Rvec}
\langle \mathbf{0}n|\hat{H}|\Rvec m\rangle ,
\label{eq:wannier_hamiltonian}
\end{equation}
where \(m,n\) label Wannier orbitals and \(\Rvec\) is a lattice vector. The same Fourier convention is used for real-space position matrix elements,
\begin{equation}
r_{nm,\alpha}^{(\mathrm{W})}(\kvec)
=
\sum_{\Rvec}
e^{\ii\kvec\cdot\Rvec}
\langle \mathbf{0}n|\hat{r}_{\alpha}|\Rvec m\rangle .
\label{eq:wannier_position_matrix}
\end{equation}
The Hamiltonian-gauge eigenstates are obtained from
\begin{equation}
H^{(\mathrm{W})}(\kvec)U(\kvec)=U(\kvec)\epsilon(\kvec),
\qquad
\epsilon_{mn}(\kvec)=\epsilon_n(\kvec)\delta_{mn},
\label{eq:eigenproblem}
\end{equation}
or, equivalently, with Hamiltonian-gauge states written without a gauge superscript,
\begin{equation}
|u_{n\kvec}\rangle
=
\sum_m |u_{m\kvec}^{(\mathrm{W})}\rangle U_{mn}(\kvec).
\label{eq:wannier_to_hamiltonian_gauge}
\end{equation}
In the manuscript notation, \(r^a_{nm}\) denotes the Hamiltonian-gauge position matrix element in Cartesian direction \(a\). It contains both the basis-derivative connection and the Wannier-gauge position matrix,
\begin{equation}
r^a_{nm}(\kvec)
\equiv
\ii\langle u_{n\kvec}|\partial_{k_a}u_{m\kvec}\rangle
=
\ii\left(U^\dagger\partial_{k_a}U\right)_{nm}
+
\left(U^\dagger r_a^{(\mathrm{W})}U\right)_{nm}.
\label{eq:hamiltonian_position_decomposition}
\end{equation}
For \(n\ne m\), the off-diagonal velocity and position matrix elements are related by
\begin{equation}
v^a_{nm}=\frac{\ii}{\hbar}\left(\epsilon_n-\epsilon_m\right)r^a_{nm},
\qquad
v^a_{nn}=\frac{1}{\hbar}\partial_{k_a}\epsilon_n .
\label{eq:velocity_position_relation}
\end{equation}
The Wannier gauge supplies the smooth, localized representation used for interpolation, whereas the Hamiltonian gauge diagonalizes \(H^{(\mathrm{W})}(\kvec)\) and is the natural basis for optical transitions. These two representations are kept together in the response kernels. WannierNLQG evaluates transition matrix elements in the Hamiltonian gauge while retaining the Wannier-gauge information required for covariant derivatives, displaced-\(\kvec\) overlaps, basis-embedding corrections, and projector derivatives.

\subsection{Symmetry-adapted Wannier construction}
\label{sec:sawf}

The symmetry-adapted construction follows the representation-constrained formulation of Sakuma \cite{Sakuma2013SAWF}. Let \(\{|u^{(\mathrm{B})}_{m\kvec}\rangle\}\) denote the Bloch states in an outer energy window. The Wannier-gauge states are obtained from the semiunitary transformation
\begin{equation}
|u^{(\mathrm{W})}_{n\kvec}\rangle
=
\sum_{m=1}^{N_{\mathrm{win}}(\kvec)}
|u^{(\mathrm{B})}_{m\kvec}\rangle V_{mn}(\kvec),
\qquad
V(\kvec)=Z(\kvec)\mathcal U(\kvec),
\label{eq:sawf_transform}
\end{equation}
where \(Z(\kvec)\) selects an \(N_{\mathrm{W}}\)-dimensional subspace from the outer window and \(\mathcal U(\kvec)\) fixes the internal Wannier gauge. Frozen-window states are retained, and the outer and frozen selections are completed to full representation blocks before the \(Z/\mathcal U\) optimization.

For \(g\) in the little group of \(\kvec\) or mapping it to \(\kvec_g\), let \(B_g(\kvec)\) be the Bloch-space sewing matrix and \(D_g(\kvec)\) the target Wannier representation. The covariance constraint is
\begin{equation}
B_g(\kvec)V(\kvec)^{*a_g}
=
V(\kvec_g)D_g(\kvec),
\qquad
a_g=
\begin{cases}
0, & g\ \text{unitary},\\
1, & g\ \text{antiunitary},
\end{cases}
\label{eq:sawf_covariance}
\end{equation}
where \(X^{*a_g}=X\) for \(a_g=0\) and \(X^{*a_g}=X^*\) for \(a_g=1\). Little-group projection enforces Eq.~\eqref{eq:sawf_covariance} at irreducible k points, and deterministic k-star expansion propagates the constrained gauge to the full mesh.

The finite-difference stencil used by the spread functional is fitted to the reciprocal metric through
\begin{equation}
\sum_b w_b\,\mathbf b\mathbf b^{\mathrm T}=I_3,
\label{eq:sawf_fd_completeness}
\end{equation}
so the real-space tight-binding export and the symmetry tests use the same lattice and Fourier conventions. The resulting Hamiltonian, position matrix, and any exported auxiliary operators form the input model for the response layer.

\subsection{Post-hoc symmetrization of Wannier operators}
\label{sec:posthoc_symmetry}

Finite numerical tolerances and approximations can leave components of an existing Wannier model that do not respect the target crystal or magnetic symmetry. Group averaging restores covariance of the hopping matrices, position matrix, and other real-space operator blocks without repeating the Wannier construction.
To describe this procedure for a general Wannier-basis operator, let \(\mathcal O^{st}(\Rvec)\) denote a block between Wannier centers \(s\) and \(t\), with local orbital and spin indices implicit. For a unitary or antiunitary operation \(g=\{W_g|\bm{\tau}_g\}\), define \(\mathbf T_{g,s}\) by \(W_g\mathbf q_s+\bm{\tau}_g=\mathbf q_{g(s)}+\mathbf T_{g,s}\), where \(\mathbf q_s\) is the fractional position of center \(s\). The mapped lattice vector is \(\Rvec_g=W_g\Rvec+\mathbf T_{g,t}-\mathbf T_{g,s}\), and \(D_{g,s}\) represents \(g\) in the local orbital--spin basis at \(s\). The symmetrized block is
\begin{equation}
\mathcal O_{\mathrm{sym}}^{st}(\Rvec)
=
\frac{1}{|G|}
\sum_{g\in G}
\mathcal T_g^{(\mathcal O)}
\!\left[
D_{g,s}^{\dagger}
\mathcal O^{g(s)g(t)}(\Rvec_g)
D_{g,t}
\right].
\label{eq:wannier_operator_symmetrization}
\end{equation}
Here \(\mathcal T_g^{(\mathcal O)}\) accounts for Cartesian rotations of physical components, polar or axial behavior under improper operations, and complex conjugation for antiunitary operations. Each symmetry-related block is pulled back to a common local basis and component frame before averaging. A Hamiltonian hopping is the Cartesian-scalar case, whereas the position matrix is a polar vector.

The diagonal on-site part of the position matrix carries the Wannier centers and is separated before group projection. With \(\tau_{s,\alpha}\) the Cartesian coordinate of center \(s\), the centerless matrix is
\begin{equation}
\widetilde r_{\alpha}^{st}(\Rvec)
=
r_{\alpha}^{st}(\Rvec)
-\delta_{\Rvec,\mathbf 0}\delta_{st}\tau_{s,\alpha}.
\label{eq:centerless_position}
\end{equation}
Equation~\eqref{eq:wannier_operator_symmetrization} is applied to \(\widetilde r_{\alpha}^{st}\), and the symmetrized centers are restored once after averaging. Missing mapped blocks are treated as zero, and the output support is extended to the symmetry-generated closure. A Wannier90 WIN file and tight-binding model supply the crystal structure, projections, local bases, and Wannier representation required for this mapping; auxiliary real-space operators can be included in the same operator bundle.

\subsection{Symmetry-reduced response integration}
\label{sec:response_symmetry}

For response integration, a unitary or antiunitary operation maps a fractional reciprocal coordinate according to
\begin{equation}
\kvec'=s_g W_g^{-\mathrm T}\kvec,
\qquad
s_g=
\begin{cases}
+1, & g\ \text{unitary},\\
-1, & g\ \text{antiunitary}.
\end{cases}
\label{eq:response_k_action}
\end{equation}
On a closed \(\Gamma\)-centered mesh, this action partitions the full set of \(N_{\mathrm{full}}\) k points into orbits. Let \(\kappa\) denote an orbit representative with multiplicity \(m_\kappa\), let \(F(\kappa,\omega)\) be its flattened response tensor, and let the real orthonormal columns \(\mathcal B_j\) of \(\mathcal B\) span the invariant tensor subspace, so that \(\mathcal B^{\mathrm T}\mathcal B=I\). The reduced accumulation is
\begin{equation}
c_j(\omega)
=
\frac{1}{N_{\mathrm{full}}}
\sum_{\kappa\in\mathrm{IBZ}}
m_\kappa\,\mathcal B_j^{\mathrm T}F(\kappa,\omega),
\qquad
X_{\mathrm{sym}}(\omega)=\mathcal Bc(\omega).
\label{eq:response_symmetry_reduction}
\end{equation}
Thus orbit weights remain normalized by the full k mesh, with \(\sum_\kappa m_\kappa=N_{\mathrm{full}}\). Because \(\mathcal B\) is real, the projector \(\mathcal B\mathcal B^{\mathrm T}\) acts separately on the real and imaginary parts of a complex response. In Cartesian form, charge-current tensors transform with \(R_g\otimes R_g\otimes R_g\), while spin current tensors use \(R_g\otimes[\det(R_g)R_g]\otimes R_g\otimes R_g\). At the stored response-tensor level, an antiunitary operation exchanges the two ordered optical-field indices, so no additional tensor conjugation is applied during projection. Combining this exchange with the time-reversal parity of charge and spin currents gives response-specific factors of \(+1\) for shift current, \(-1\) for injection current, \(-1\) for shift spin current, and \(+1\) for injection spin current \cite{Sipe2000SecondOrder,Xu2021BulkSpinPhotovoltaic}. These response transformations determine \(\mathcal B\), while the magnetic symmetry operations used to construct them are obtained with Spglib \cite{Togo2024Spglib,Shinohara2023MagneticSymmetry}.

\subsection{Second-order dc response: shift and injection currents}
\label{sec:shift_current}

For a monochromatic optical field, the dc component of a spatially uniform second-order photocurrent is written as
\begin{equation}
j^a_{\mathrm{dc}}
=
2\sum_{bc}
\left[
\sigma^{abc}_{\mathrm{IC}}(\omega)
+
\sigma^{abc}_{\mathrm{SC}}(\omega)
\right]
E^b(\omega)E^c(-\omega).
\end{equation}
The field convention is \(\mathbf{E}(t)=\mathbf{E}(\omega)e^{-\ii\omega t}+\mathbf{E}(-\omega)e^{\ii\omega t}\). The prefactor \(2\) collects the conjugate frequency orderings \((\omega,-\omega)\) and \((-\omega,\omega)\), and each \(\sigma^{abc}(\omega)\) denotes the ordered-field contribution associated with \(E^b(\omega)E^c(-\omega)\). The sum over \(b,c\) retains ordered Cartesian field indices. The two tensors, \(\sigma^{abc}_{\mathrm{IC}}(\omega)\) and \(\sigma^{abc}_{\mathrm{SC}}(\omega)\), denote the injection current and shift current photoconductivities. In the clean limit, where the dc response is dominated by resonant interband transitions across the optical gap, these are the leading spatially uniform mechanisms: injection current is controlled by the velocity imbalance of optically injected carriers, whereas shift current is controlled by the coordinate displacement associated with an optical transition.

The steady injection-current photoconductivity can be written as \cite{Sipe2000SecondOrder,Wang2017NLOWannier}
\begin{equation}
\sigma^{abc}_{\mathrm{IC}}(\omega)=
\frac{\tau\pi e^3}{\hbar^2}
\int_{\mathrm{BZ}}\frac{\dk}{(2\pi)^d}
\sum_{nm} f_{nm}\,
\delta(\omega_{mn}-\omega)\,
\left(v^a_{mm}-v^a_{nn}\right)
r^b_{mn}r^c_{nm}.
\label{eq:injection_current}
\end{equation}
Here \(\tau\) is the relaxation time used when a steady injection current is reported. The factor \(v^a_{mm}-v^a_{nn}\) is the group-velocity imbalance between the optically connected bands, and the theoretical delta function is represented numerically by the selected broadening function.

Shift current has a different physical origin. It is generated by the real-space displacement of an electronic wave packet during an interband optical transition, and therefore does not require a relaxation-time imbalance.

For a spatially uniform optical field, the conventional expression can be written in the form
\begin{equation}
\sigma^{abc}_{\mathrm{SC}}(\omega)=
\frac{\pi e^3}{2\hbar^2}
\int_{\mathrm{BZ}}\frac{\dk}{(2\pi)^d}
\sum_{n,m} f_{nm}\,
\delta(\omega_{mn}-\omega)
r^b_{mn}r^c_{nm}
\left(
R_{mn}^{b;a}
-R_{nm}^{c;a}
\right),
\label{eq:shift_current_general}
\end{equation}
where \(f_{nm}=f_n-f_m\), \(\omega_{mn}=(\epsilon_m-\epsilon_n)/\hbar\), \(e=-|e|<0\) denotes the electron charge, and \(a,b,c\) are Cartesian tensor indices. The shift vector is
\begin{equation}
R_{mn}^{b;a}
=
\ii\partial_{k^a}\log r^b_{mn}
+r^a_{mm}
-r^a_{nn},
\label{eq:shift_vector}
\end{equation}
Equations~\eqref{eq:shift_current_general} and \eqref{eq:shift_vector} define the standard shift vector form of the conventional shift current response \cite{Sipe2000SecondOrder,Wang2017NLOWannier,IbanezAzpiroz2018ShiftWannier}. In general, \(R_{mn}^{b;a}\) is a generalized complex shift vector building block: where \(r^b_{mn}\neq0\), its real part gives the transition displacement from band \(n\) to band \(m\), while its imaginary part contains the derivative of the transition-amplitude magnitude. At zeros of \(r^b_{mn}\), the logarithmic expression itself is undefined, but the conductivity uses the optically weighted combination in Eq.~\eqref{eq:shift_current_general}, evaluated numerically through finite sum-rule intermediates. The dc current is the Brillouin-zone sum of these optically driven charge displacements.

Equations~\eqref{eq:shift_current_general} and \eqref{eq:shift_vector} are not evaluated directly in the conventional numerical route. The logarithmic derivative in Eq.~\eqref{eq:shift_vector} contains the phase of the interband matrix element, and direct differentiation of this gauge-dependent quantity is not stable in numerical work. The conventional implementation therefore evaluates the same shift vector building block through the sum rule for the generalized derivative. In the present notation, one may write
\begin{equation}
r^b_{mn} R^{b;a}_{mn}
=
\frac{1}{\omega_{mn}}
\left[
-
\frac{
v^b_{mn}\Delta^a_{mn}
+
v^a_{mn}\Delta^b_{mn}
}{\omega_{mn}}
+
w^{ba}_{mn}
-
\sum_{p\neq m,n}
\left(
\frac{v^b_{mp}v^a_{pn}}{\omega_{pn}}
-
\frac{v^a_{mp}v^b_{pn}}{\omega_{mp}}
\right)
\right].
\label{eq:shift_sum_rule}
\end{equation}
Here \(\Delta^a_{mn}=v^a_{mm}-v^a_{nn}\), and \(w^{ba}_{mn}\) denotes the Hamiltonian second-derivative matrix element \(\hbar^{-1}\langle u_m|\partial_{k_b}\partial_{k_a}\hat{H}|u_n\rangle\). Equation~\eqref{eq:shift_sum_rule} applies to interband pairs with \(m\neq n\). It replaces the phase derivative in Eq.~\eqref{eq:shift_vector} with velocity matrix elements, energy denominators, and a second derivative of the Hamiltonian, all evaluated at the same \(\kvec\). This is the numerical advantage of the sum-rule form.

The remaining cost in Eq.~\eqref{eq:shift_sum_rule} comes from the explicit sum over intermediate states \(p\). In a direct first-principles implementation, convergence may require many unoccupied bands, and a finite band cutoff introduces band-truncation errors in the generalized derivative. Wannier interpolation formulations of nonlinear optical response and shift photocurrent avoid this Bloch-band truncation problem by rewriting the generalized derivative inside the finite subspace spanned by the wannierized bands \cite{Wang2017NLOWannier,IbanezAzpiroz2018ShiftWannier}. The intermediate-state sums then run over Wannier states rather than over the full set of Bloch eigenstates, keeping the calculation compact while preserving the matrix element information needed for the shift current kernel.

The conventional route used here inherits this computational advantage. Compared with a direct first-principles evaluation, Wannier interpolation makes the cost low enough for dense Brillouin-zone sampling and does not require an explicit sum over all Bloch bands. A separate limitation remains: the formula is built from single-band interband pairs and assumes \(m\neq n\). The corresponding numerical implementation treats \(m=n\) terms and vanishing energy denominators by setting the singular single-band contribution to zero or by regularizing the inverse frequency denominator, for example as \(\omega_{np}^{-1}\rightarrow \omega_{np}/[\omega_{np}^2+(\eta/\hbar)^2]\). This prescription is practical, but it cannot represent the finite gauge-covariant response that may reside in a degenerate subspace. The projector formulation below addresses that issue by using subspace projectors rather than individual eigenvectors as the elementary objects.

\subsection{Projector formulation for shift current}
\label{sec:projector_formulation}

The final shift current tensor remains gauge invariant; the numerical limitation arises from using individual eigenvectors as intermediate objects. At a degeneracy, arbitrary unitary rotations within the degenerate manifold make single-band derivatives singular or basis dependent. The projector formulation replaces these basis-dependent intermediates with projectors. For an isolated band \(m\), the band projector is
\begin{equation}
P_m(\kvec)=|u_m(\kvec)\rangle\langle u_m(\kvec)|.
\label{eq:band_projector}
\end{equation}

Projector-based formulations express nonlinear optical response and objects in quantum geometry as traces of projectors and their momentum derivatives \cite{Avdoshkin2025MultistateGeometry,Guo2025Projector}. The QHC building block entering the projector shift current expression is
\begin{equation}
C^{a;bc}_{nm} =
\Tr\!\left[
P_m(\partial_{k_b}P_n)
\left(
\partial_{k_a}\partial_{k_c}P_m
+(\partial_{k_a}P_n)(\partial_{k_c}P_m)
\right)
\right],
\label{eq:projector_c_block}
\end{equation}
where \(m,n\) label bands and \(a,b,c\) are Cartesian tensor indices. With the source convention adapted to the notation of this manuscript, the projector shift current photoconductivity tensor is
\begin{equation}
\sigma^{abc}_{\mathrm{SC}}(\omega)=
\frac{-\ii\pi e^3}{2\hbar^2}
\int_{\mathrm{BZ}}\frac{\dk}{(2\pi)^d}
\sum_{n,m} f_{nm}\,
\delta(\omega_{mn}-\omega)
\left(
C^{a;cb}_{mn}
-C^{a;bc}_{nm}
\right).
\label{eq:projector_shift_current}
\end{equation}
For isolated bands, the projector block is equivalent to the shift vector building block in the conventional formula, \(C^{a;bc}_{nm}=\ii r^b_{mn}r^c_{nm}R_{nm}^{c;a}\). Exchanging \(n\leftrightarrow m\) and \(b\leftrightarrow c\) gives \(C^{a;cb}_{mn}=\ii r^b_{mn}r^c_{nm}R_{mn}^{b;a}\). Substitution into Eq.~\eqref{eq:projector_shift_current} yields \(\ii r^b_{mn}r^c_{nm}(R_{mn}^{b;a}-R_{nm}^{c;a})\), which, with the prefactor in Eq.~\eqref{eq:projector_shift_current}, reproduces Eq.~\eqref{eq:shift_current_general}. Thus the projector expression is a trace-form rewriting of the conventional shift current formula for isolated bands, while also providing a notation that can be promoted directly to subspaces.

When several bands form a degenerate manifold, individual eigenvectors within that manifold are not unique because any unitary rotation inside the manifold gives an equivalent basis. The gauge-covariant replacement is to promote the isolated-band projector to the projector onto the selected subspace \(N\),
\begin{equation}
P_n \equiv |u_n(\kvec)\rangle\langle u_n(\kvec)|
\;\longrightarrow\;
P_N(\kvec)=\sum_{j\in N}|u_j(\kvec)\rangle\langle u_j(\kvec)|.
\label{eq:subspace_projector}
\end{equation}
The subspace projector is invariant under basis rotations within \(N\). Response expressions written as projector traces or closed loops therefore remain well defined even when the individual band labels inside \(N\) are not.

This subspace principle is shared by the gauge-covariant routes used here. In the projector formulation, single-band projectors in objects such as Eq.~\eqref{eq:projector_c_block} are replaced by subspace projectors. In generalized Wilson loop and finite-\(\qvec\) geometric loop formulations, scalar links associated with individual bands are replaced by block contractions between the relevant valence and conduction subspaces, while the full loop remains the gauge-invariant numerical object. The QHC k-slice quantity implemented in WannierNLQG uses the same subspace treatment and serves here as a diagnostic of this general degeneracy-handling strategy.

\subsection{Nonlinear photon drag currents in the conventional formalism}
\label{sec:pdsc_conventional}
\label{sec:pdic}

Spatially uniform second-order dc photocurrents are strongly constrained by crystal symmetry and can vanish in inversion-symmetric systems or in selected tensor channels. Responses driven by photon drag retain the finite photon momentum as an additional polar vector. This nonzero \(\qvec\) relaxes the \(q=0\) symmetry restrictions and opens nonlinear current channels that are inaccessible in the spatially uniform electric dipole limit \cite{Shi2021PhotonDrag,Xie2025PhotonDragBPVE,Lu2025GiantPhotonDrag,Zhao2025MoS2CircularPhotocurrent}.

For a monochromatic optical field carrying a finite photon wave vector, the dc photon drag response considered here is written as
\begin{equation}
j^a_{\mathrm{dc}}(\qvec)
=
2\sum_{bc}
\left[
\sigma^{abc}_{\mathrm{PDIC}}(\qvec,\omega)
+
\sigma^{abc}_{\mathrm{PDSC}}(\qvec,\omega)
\right]
E^b(\qvec,\omega)E^c(-\qvec,-\omega).
\end{equation}
The field convention is \(\mathbf{E}(\mathbf{r},t)=\mathbf{E}(\qvec,\omega)e^{\ii\qvec\cdot\mathbf{r}-\ii\omega t}+\mathrm{c.c.}\). The complex-conjugate term corresponds to \(\mathbf{E}(-\qvec,-\omega)e^{-\ii\qvec\cdot\mathbf{r}+\ii\omega t}\). The prefactor \(2\) collects the conjugate frequency--momentum orderings \((\qvec,\omega;-\qvec,-\omega)\) and \((-\qvec,-\omega;\qvec,\omega)\), and each finite-\(\qvec\) conductivity denotes the ordered-field contribution multiplying \(E^b(\qvec,\omega)E^c(-\qvec,-\omega)\). The sum over \(b,c\) retains ordered Cartesian field indices. The \(q\)-odd post-processing is applied separately. These conductivities are the finite-\(\qvec\) counterparts of the spatially uniform injection and shift currents. Photon momentum connects displaced crystal momenta \(\kvec_\alpha=\kvec-\qvec/2\) and \(\kvec_\beta=\kvec+\qvec/2\) in the optical transition.

The photon drag injection current (PDIC) response is the finite-\(\qvec\) analogue of injection current and is controlled by the velocity imbalance of the optically connected states. In the same finite-\(\qvec\) convention, the conventional formula is \cite{Lu2025GiantPhotonDrag}
\begin{equation}
\sigma^{abc}_{\mathrm{PDIC}}(\qvec,\omega)=
\frac{\tau\pi e^3}{\hbar^2\omega^2}
\int_{\mathrm{BZ}}\frac{\dk}{(2\pi)^d}
\sum_{n_\alpha,n_\beta}
f_{\alpha\beta}
\delta(\omega_{\beta\alpha}-\omega)
\left(v^a_{\beta\beta}-v^a_{\alpha\alpha}\right)
v_{\beta\alpha}^{b}v_{\alpha\beta}^{c}.
\label{eq:pdic}
\end{equation}
Here \(\tau\) is the relaxation time used when a steady injection current is reported. Finite photon momentum enters through the transition vertices \(v_{\beta\alpha}^{b}\) and \(v_{\alpha\beta}^{c}\), which connect \(\kvec_\alpha\) and \(\kvec_\beta\). In the implementation, these optical vertices are constructed from overlaps between displaced Hamiltonian-gauge eigenvectors and an averaged velocity operator, while the velocity imbalance \(v^a_{\beta\beta}-v^a_{\alpha\alpha}\) enters the kernel directly.

The photon drag shift current (PDSC) response is instead controlled by a transition displacement at finite momentum. The conventional shift vector formula follows the convention of the photon drag manuscript \cite{Shi2021PhotonDrag,Xie2025PhotonDragBPVE,Lu2025GiantPhotonDrag},
\begin{equation}
\sigma^{abc}_{\mathrm{PDSC}}(\qvec,\omega)=
\frac{\pi e^3}{2\hbar^2\omega^2}
\int_{\mathrm{BZ}}\frac{\dk}{(2\pi)^d}
\sum_{n_\alpha,n_\beta}
f_{\alpha\beta}
\delta(\omega_{\beta\alpha}-\omega)
\left(
R_{\beta\alpha}^{b;a}
-R_{\alpha\beta}^{c;a}
\right)
v_{\beta\alpha}^{b}v_{\alpha\beta}^{c},
\label{eq:pdsc_shift_vector}
\end{equation}
where
\begin{equation}
R_{\beta\alpha}^{b;a}
=
\ii\partial_{k_a}\log v_{\beta\alpha}^{b}
+r_{\beta\beta}^{a}
-r_{\alpha\alpha}^{a}.
\label{eq:pdsc_shift_vector_def}
\end{equation}
Here the composite indices at finite momentum are \(\alpha=(n_\alpha,\kvec_\alpha)\) and \(\beta=(n_\beta,\kvec_\beta)\), with \(\kvec_\alpha=\kvec-\qvec/2\) and \(\kvec_\beta=\kvec+\qvec/2\). The resonant factor in Eq.~\eqref{eq:pdsc_shift_vector} is \(f_{\alpha\beta}\delta(\omega_{\beta\alpha}-\omega)\), where \(f_{\alpha\beta}=f_\alpha-f_\beta\), \(\omega_{\beta\alpha}=(\epsilon_\beta-\epsilon_\alpha)/\hbar\), and \(|u_\alpha\rangle\equiv |u_{n_\alpha\kvec_\alpha}\rangle\). The finite-\(\qvec\) optical velocity vertex is \(v_{\alpha\beta}^{b}=\langle u_{\alpha}|\hat{v}_{\alpha,\beta}^{b}|u_{\beta}\rangle\), with \(\hat{v}_{\alpha,\beta}^{b}=(\hat{v}_{\kvec_\alpha}^{b}+\hat{v}_{\kvec_\beta}^{b})/2\); the exchanged-index vertex is defined analogously. The quantity \(R_{\beta\alpha}^{b;a}\) at finite momentum is likewise generally complex, and its relevant real part gives the transition displacement wherever \(v_{\beta\alpha}^{b}\neq0\). At a zero of the optical vertex, the logarithmic expression is undefined, whereas the weighted kernel in Eq.~\eqref{eq:pdsc_shift_vector} remains the physical object and is evaluated here through the geometric loop. The PDSC kernel is therefore weighted by a shift vector at finite momentum and by optical vertices connecting \(\kvec_\alpha\) and \(\kvec_\beta\).

The conventional PDSC shift vector form has the same numerical weakness as the spatially uniform shift current expression. The derivative in Eq.~\eqref{eq:pdsc_shift_vector_def} acts on the phase of the finite-\(\qvec\) velocity matrix element, which is gauge dependent, so direct numerical differentiation is fragile. Sum-rule forms can avoid this derivative, but they involve intermediate-state sums and inherit the same single-band limitations at degeneracies as the \(q=0\) conventional formula. This motivates the geometric loop formulation below, where the same finite-\(\qvec\) shift current kernel is evaluated from a closed gauge-invariant loop.

\subsection{Geometric-loop formulation for photon drag shift current}
\label{sec:pdsc_geometric_loop_formulation}

For numerical stability, WannierNLQG evaluates PDSC with the geometric loop form. Define
\begin{equation}
\mathcal{L}_{\beta\alpha}^{abc}(p)=
\langle
P_{\beta'}\hat{v}^{b}_{\beta'\alpha'}P_{\alpha'}
P_{\alpha}\hat{v}^{c}_{\alpha\beta}P_{\beta}
\rangle_{\beta},
\label{eq:geometric_loop_pd}
\end{equation}
where \(\langle X\rangle_{\beta}\equiv \langle u_{\beta}|X|u_{\beta}\rangle\), \(\alpha'=(n_\alpha,\kvec_\alpha+p\hat{\mathbf{e}}_a)\), \(\beta'=(n_\beta,\kvec_\beta+p\hat{\mathbf{e}}_a)\), and the limit \(p\to0^+\) is taken after differentiation. With \(\hat{D}_p=\ii\partial_p\),
\begin{equation}
\sigma^{abc}_{\mathrm{PDSC}}(\qvec,\omega)=
\frac{\pi e^3}{2\hbar^2\omega^2}
\int_{\mathrm{BZ}}\frac{\dk}{(2\pi)^d}
\sum_{n_\alpha,n_\beta}
f_{\alpha\beta}
\delta(\omega_{\beta\alpha}-\omega)
\left[
\hat{D}_p\mathcal{L}_{\beta\alpha}^{abc}(p)
-\hat{D}_p\mathcal{L}_{\alpha\beta}^{acb}(p)
\right]_{p\to0^+}.
\label{eq:pdsc_geometric_loop}
\end{equation}
This representation makes the connection between the finite-\(\qvec\) shift vector and the geometric loop explicit. In the \(p\to0^+\) limit, the loop amplitude reduces to the optical-velocity product, \(v_{\beta\alpha}^{b}v_{\alpha\beta}^{c}=\mathcal{L}_{\beta\alpha}^{abc}(p)\), while the shift vector is obtained from the logarithmic loop derivative, \(R_{\beta\alpha}^{b;a}=\hat{D}_p\log\mathcal{L}_{\beta\alpha}^{abc}(p)\). Equivalently, the product \(v_{\beta\alpha}^{b}v_{\alpha\beta}^{c}R_{\beta\alpha}^{b;a}\) in Eq.~\eqref{eq:pdsc_shift_vector} is represented by \(\hat{D}_p\mathcal{L}_{\beta\alpha}^{abc}(p)\), with the exchanged loop giving the \(R_{\alpha\beta}^{c;a}\) term. Because the loop is a closed product of projectors and finite-\(\qvec\) velocity vertices connecting \(\kvec_\alpha\) and \(\kvec_\beta\), local gauge phases cancel within the complete object. The PDSC response can then be evaluated by finite differences of a gauge-invariant loop rather than by differentiating gauge-dependent velocity matrices or using multi-band sum-rule expressions.

\section{Software implementation}
\label{sec:software}

\subsection{Architecture and public interface}

WannierNLQG v1.0.0 exposes a grouped public response interface. A \code{TaskConfig} combines a \code{ModelInput}, one Brillouin-zone sampling object, named \code{TaskSpec} objects, and execution and output settings. Task-specific parameter and observable types complete the configuration; \code{run(cfg)} returns a \code{RunResult}. The \code{Core} and \code{IO} layers supply model data, \code{MatrixElements} constructs the required geometric and optical objects, \code{Responses} evaluates observables, and \code{Runtime} orchestrates validation, execution, and output. Shared symmetry and projection infrastructure supports the expert \code{Wannierization} and \code{Symmetrization} namespaces. A central task registry defines the supported quantity--method--calculation combinations. After activating the package environment, users adapt a maintained example from \code{examples/tasks/} or write a Julia calculation script; the same interface serves local calculations and cluster jobs. Setup rejects incompatible task bundles and missing input capabilities before entering the k loop.

Fig.~\ref{fig:workflow} summarizes the response workflow. Tasks share model preparation and spectral data, while dependency-based planning constructs only the matrix elements, projectors, spin operators, or loop objects needed by the bundle. The fused k loop uses worker-local workspaces and accumulators with deterministic reduction. Threading and optional MPI distribute k points; the root process writes the final outputs. Ordinary local runs do not initialize MPI unless it is explicitly enabled.

\subsection{Task configuration and supported calculations}

The \code{tasks} field accepts named \code{TaskSpec} objects with unique identifiers; the method can be omitted when uniquely determined, and the calculation type is inferred from the shared \code{BZMesh}, \code{KSlice}, or \code{KPath} sampling object. Table~\ref{tab:integral_tasks} summarizes the supported response calculations in both Integral and K-slice modes. Table~\ref{tab:kslice_tasks} lists the quantities in quantum geometry available for K-slice calculations. Short labels and full quantity names are interchangeable within their registered calculation type. An additional, independent \code{Band} task evaluates eigenvalues along a specified k path using Direct Fourier interpolation and runs as a single-task bundle. A \code{TaskConfig} shares its model, sampling, execution, and output settings, while each \code{TaskSpec} owns its physical parameters, numerical controls, and observable. Integral tasks use \code{FullTensor()}, whereas K-slice tasks use \code{KSliceSelection} to select a tensor component and explicit bands or subspaces. Optical frequencies, occupations, broadening, finite-difference steps, and \code{degeneracy_threshold} are specified for the tasks to which they apply. Tasks at finite momentum use \code{FiniteQOpticalParameters} to supply \code{photon_momentum}. The conventional formulation uses single-band intermediates; projector and loop-block formulations supply the corresponding degenerate-subspace treatment.

Listing~\ref{lst:input} illustrates the grouped interface with the synthetic inputs shipped in the package. The user guide and maintained task examples document the parameter groups, task-specific inputs, and registered labels.

\begin{lstlisting}[caption={Two-formulation synthetic integral configuration for WannierNLQG v1.0.0.},label={lst:input}]
using WannierNLQG
import WannierNLQG: run

fixture_root = joinpath(pkgdir(WannierNLQG),
    "examples", "fixtures", "synthetic_runtime")
physics = OpticalParameters(
    photon_energies = collect(range(0.0, 4.0; length=200)),
    fermi_energy = 0.0, temperature = 0.0,
)
base_numerics = (
    broadening = 0.060, broadening_type = "Gaussian",
    transition_window_factor = 5.0, denominator_regularization = 0.001,
    band_window_size = -1,
)

cfg = TaskConfig(
    model = ModelInput(
        case_root = fixture_root,
        model_file = "synthetic_tb.dat",
        real_space_operator_bundle_file = "synthetic_operators.h5",
        real_space_replica_policy = "minimum_distance",
        wsvec_file = "synthetic_wsvec.dat", mp_grid = (2, 1, 1),
        wannier_center_convention = "Convention_II",
    ),
    sampling = BZMesh(k_mesh = (100, 100), spatial_dimension = 2),
    tasks = [
        TaskSpec(
            id = "conventional", quantity = "SC", method = "Conventional",
            physics = physics, observable = FullTensor(),
            numerics = OpticalNumerics(; base_numerics...),
        ),
        TaskSpec(
            id = "projector", quantity = "SC", method = "Projector",
            physics = physics, observable = FullTensor(),
            numerics = OpticalNumerics(
                ; base_numerics...,
                finite_difference_step = 0.0001,
                degeneracy_threshold = 2.0e-3,
            ),
        ),
    ],
    execution = ExecutionOptions(fourier_backend = "direct"),
    output = OutputOptions(
        output_root = joinpath(
            @__DIR__, "outputs", "synthetic_shift_current"),
        system_name = "synthetic_demo",
    ),
)
result = run(cfg)
\end{lstlisting}

The required \code{fourier_backend} selects \code{"direct"}, \code{"mixed"}, or \code{"auto"}. Mixed interpolation combines coarse blocks and local FFTs using \code{NKdiv} and \code{NKFFT}; one factor can be inferred by exact division. Auto uses Direct without factors and otherwise evaluates the supplied Mixed plan according to feasibility, memory, and the estimated symmetry-reduced workload where applicable. Band paths use Direct interpolation. The \code{wannier_center_convention} selects the Fourier convention, with physical position and external terms included consistently.

\begin{table}[!htbp]
\centering
\caption{Supported response calculations in WannierNLQG v1.0.0. Each quantity supports both Integral and K-slice calculations with the methods listed. Short labels are given in Integral / K-slice order.}
\label{tab:integral_tasks}
\small
\begin{tabular}{@{}>{\raggedright\arraybackslash}p{0.42\linewidth}>{\raggedright\arraybackslash}p{0.46\linewidth}@{}}
\toprule
Quantity & Supported methods \\
\midrule
\code{Shift_Current} (\code{SC} / \code{SCK}) & \code{Conventional}, \code{Projector}, \code{Geometric_Loop}, \code{Wilson_Loop} \\
\code{Injection_Current} (\code{IC} / \code{ICK}) & \code{Conventional} \\
\code{Injection_Spin_Current} (\code{ISC} / \code{ISCK}) & \code{Conventional} \\
\code{Shift_Spin_Current} (\code{SSC} / \code{SSCK}) & \code{Conventional} \\
\code{Photon_Drag_Injection_Current} (\code{PDIC} / \code{PDICK}) & \code{Conventional} \\
\code{Photon_Drag_Shift_Current} (\code{PDSC} / \code{PDSCK}) & \code{Geometric_Loop} \\
\bottomrule
\end{tabular}
\end{table}

\begin{table}[!htbp]
\centering
\caption{Supported K-slice calculations of quantities in quantum geometry in WannierNLQG v1.0.0.}
\label{tab:kslice_tasks}
\small
\begin{tabular}{@{}>{\raggedright\arraybackslash}p{0.62\linewidth}>{\raggedright\arraybackslash}p{0.30\linewidth}@{}}
\toprule
Quantity & Supported methods \\
\midrule
\code{Shift_Vector} (\code{SVK}) & \code{Geometric_Loop}, \code{Wilson_Loop} \\
\code{Quantum_Hermitian_Connection} (\code{QHCK}) & \code{Conventional}, \code{Projector}, \code{Geometric_Loop}, \code{Wilson_Loop} \\
\code{Quantum_Christoffel_Symbol} (\code{QCSK}) & \code{Conventional} \\
\code{Hermitian_Curvature_Tensor} (\code{HCTK}) & \code{Conventional} \\
\code{Triple_Phase_Product} (\code{TPPK}) & \code{Conventional} \\
\code{Berry_Curvature} (\code{BCK}), \code{Quantum_Metric} (\code{QMK}) & \code{Conventional} \\
\code{Interband_Berry_Curvature} (\code{IBCK}), \code{Interband_Quantum_Metric} (\code{IQMK}) & \code{Conventional} \\
\code{Berry_Curvature_Dipole} (\code{BCDK}), \code{Quantum_Metric_Dipole} (\code{QMDK}) & \code{Conventional} \\
\code{Berry_Curvature_Quadrupole} (\code{BCQK}), \code{Quantum_Metric_Quadrupole} (\code{QMQK}) & \code{Conventional} \\
\code{Zeeman_Interband_Berry_Curvature} (\code{ZIBCK}), \code{Zeeman_Interband_Quantum_Metric} (\code{ZIQMK}) & \code{Conventional} \\
\bottomrule
\end{tabular}
\end{table}

\FloatBarrier

\subsection{Model preparation and symmetry workflows}

Models enter through a Wannier tight-binding text file or a self-contained HDF5 real-space operator bundle selected by \code{real_space_operator_bundle_file}. Preparation resolves the real-space replica mapping before Fourier interpolation, using explicit Wannier90 replica data or a reconstruction from the mesh and Wannier centers; already materialized mappings are reused without duplication. The \code{real_space_replica_policy} records whether the input support or minimum-distance replicas are used. With the default Auto policy, a text model lacking replica information retains its input support, and that choice is recorded. Each task checks the required operator inventory. For legacy inputs, spin current tasks use a common \code{seedname}, while Zeeman-geometry tasks load matching spin and checkpoint files; HDF5 inputs instead supply the corresponding spin and mixed operators. Projector shift current and QHC tasks require a Packed HDF5 bundle containing the complete exact-\code{uIu} derivative-overlap tensor and its source provenance.

The expert \nolinkurl{WannierNLQG.Wannierization} namespace constructs Wannier models from VASP or QE inputs, including symmetry-adapted construction. Its configuration groups input, solver, checkpoint, runtime, and output settings. \nolinkurl{WannierNLQG.Symmetrization} projects existing Hamiltonian and position matrices and supported auxiliary operators onto a target symmetry group. SAWF construction uses serial or threaded execution. These preprocessing workflows produce models, checkpoints, operator bundles, and symmetry records and can be used independently of response integration.

A matching \code{response_symmetry_file} enables symmetry analysis of supported \(q=0\) charge- and spin current tasks. In reduced Integral mode, irreducible k-point orbits on a closed \(\Gamma\)-centered mesh carry full-mesh-normalized weights, followed by invariant-tensor reconstruction. Full-mesh Integral mode can instead produce a symmetry report without projecting the tensor. K-slice analysis reports component relations without reducing or projecting the slice data; finite-\(q\) responses retain their unreduced workflow. The fields \code{response_symmetry_kmesh_mode} and \code{response_symmetry_report_enabled} select the mesh and optional full-mode report. With no artifact, the default response calculation uses the full mesh. The default strict policy validates model identity and artifact eligibility before use. Task-specific method and input requirements are checked during setup.

\subsection{Outputs and reproducibility}

Each run uses \code{output_root} as its output parent and writes every task under \code{output_root/<id>/}. Integral files contain Fermi energy, photon energy, and Cartesian tensor components; K-slice files contain the selected local quantities, with separate real and imaginary files where needed. Band calculations produce eigenvalues and path metadata. Filenames identify the system, observable, and formulation. The returned \code{RunResult} collects normalized task specifications, aggregate output paths, task identifiers, per-task results, and sharing statistics.

Each task's \code{metadata.txt} records model identity, numerical settings, band or subspace selections, the resolved Fourier and replica plans, and output files. The root \code{metadata.txt} indexes the task identifiers, child metadata, outputs, and sharing statistics. Optional \code{WannierNLQG.out} and \code{progress.jsonl} provide human-readable progress and machine-readable events. Symmetry reports and model-preparation records connect the response to its input model and symmetry assumptions. Together, these records support reproducibility across local and distributed runs; the user guide specifies the complete file formats.

\section{Examples and validation}
\label{sec:examples}

\subsection{GeS: shift current and quantum geometry}

Monolayer GeS is a representative two-dimensional group-IV monochalcogenide predicted to exhibit a large bulk photovoltaic effect and spontaneous polarization \cite{Rangel2017GeS}. Its shift current response has been studied extensively in first-principles calculations using Wannier interpolation, the generalized Wilson loop formalism, and the projector method \cite{Wang2017NLOWannier,IbanezAzpiroz2018ShiftWannier,Wang2022WilsonLoop,Guo2025Projector}. We use GeS to compare the DFT and Wannier-TB electronic structures, assess the agreement among the conventional, projector, and generalized Wilson loop shift current formulations under matched settings, and resolve the k-dependent Wilson loop QHC and shift vector for near-edge valence and conduction subspaces.

\begin{figure*}[!tbp]
\centering
\includegraphics[width=0.92\textwidth,trim=0 30bp 0 44bp,clip]{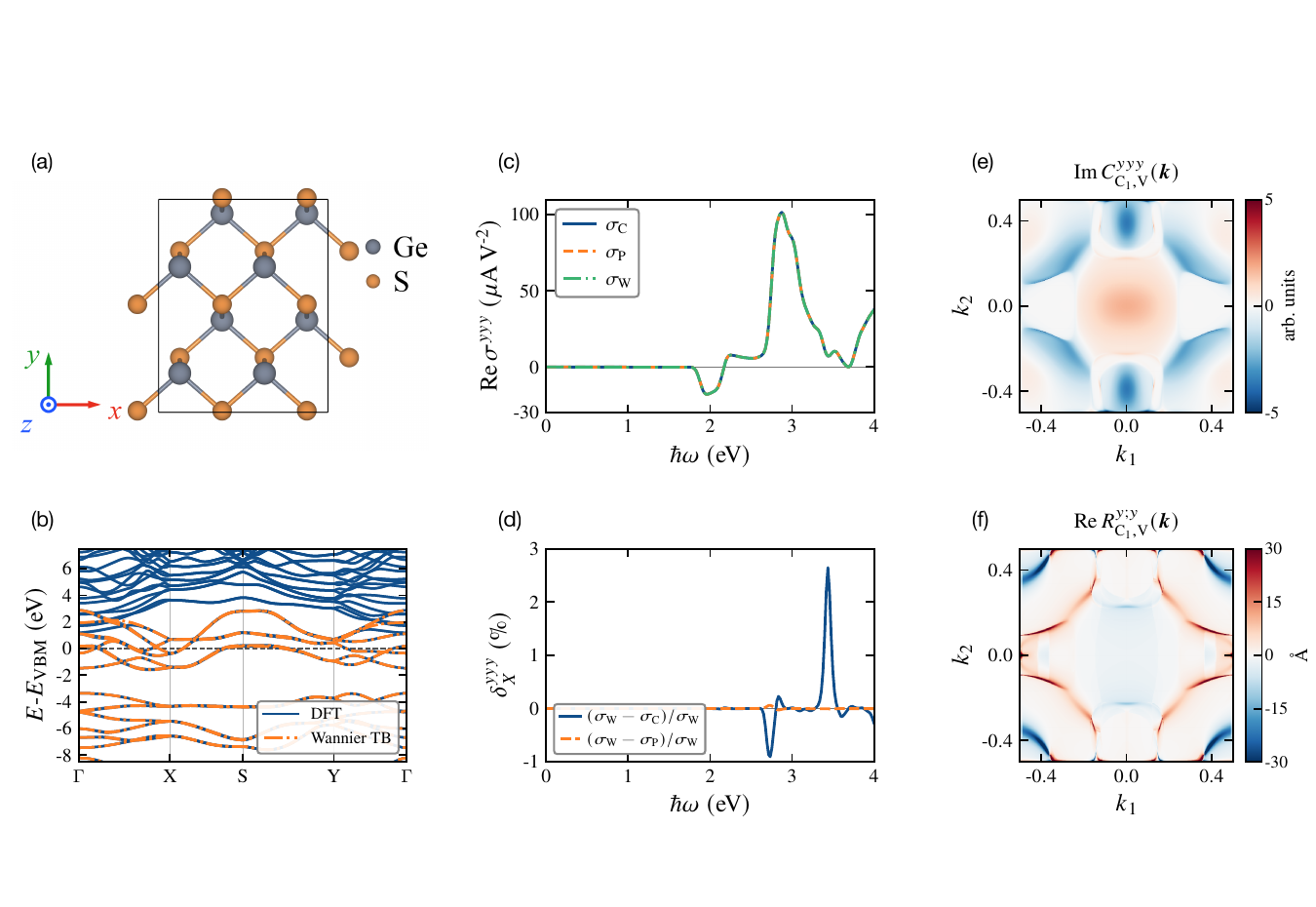}
\caption{GeS electronic structure, shift current comparison, and Wilson loop k-space quantities, arranged as \((a,c,e)\) in the upper row and \((b,d,f)\) in the lower row. (a) POSCAR-derived lattice schematic. (b) DFT and Wannier-TB band structures along \(\Gamma\)--X--S--Y--\(\Gamma\), referenced to the valence-band maximum. (c) Integrated \(\mathrm{Re}\,\sigma^{yyy}\) spectra from the conventional, projector, and Wilson loop formulations, denoted by \(\sigma_{\mathrm{C}}\), \(\sigma_{\mathrm{P}}\), and \(\sigma_{\mathrm{W}}\), respectively. The spectra are calculated under matched numerical settings using Wannier-center Convention II, on a \(200\times200\times1\) mesh and a 200-point \(0\)--\(4~\mathrm{eV}\) frequency grid, and are displayed as effective-bulk conductivities in \(\mu\mathrm{A}\,\mathrm{V}^{-2}\) using the thickness factor \(L_z/d_{\mathrm{eff}}=(25.4~\mathrm{\mathring{A}})/(2.6~\mathrm{\mathring{A}})\). (d) Local percentage differences relative to the Wilson loop result, labeled \(\delta_{\mathrm C}^{yyy}\) and \(\delta_{\mathrm P}^{yyy}\). The plotted curves are evaluated as \(\delta_X^{yyy}=100(\sigma_{\mathrm{W}}^{yyy}-\sigma_X^{yyy})/D_{\mathrm{W}}\), where \(D_{\mathrm{W}}(\omega)=\max\{|\sigma_{\mathrm{W}}^{yyy}(\omega)|,0.05\sigma_{\mathrm{pk}}\}\) and \(\sigma_{\mathrm{pk}}=\max_{\omega'}|\sigma_{\mathrm{W}}^{yyy}(\omega')|\). Thus the denominator follows the magnitude of the Wilson loop result where it is resolved and uses a small peak-scaled floor to remain finite as the reference approaches zero. (e) Wilson loop \(\mathrm{Im}\,C^{yyy}_{\mathrm{C}_1,\mathrm{V}}(\mathbf{k})\), shown over the symmetric range \([-5,5]\). (f) Wilson loop \(\mathrm{Re}\,R^{y;y}_{\mathrm{C}_1,\mathrm{V}}(\mathbf{k})\), shown over \([-30,30]~\mathrm{\mathring{A}}\) with a linear diverging color scale. For readability, these display ranges clip approximately \(0.32\%\) and \(1.72\%\) of the grid points in panels (e) and (f), respectively; the underlying arrays retain their full values. Panels (e) and (f) use \(\mathrm{C}_1=[21,22]\), \(\mathrm{V}=[19,20]\), and the same centered \(400\times400\times1\) k-slice convention.}
\label{fig:ges_suite}
\end{figure*}

Fig.~\ref{fig:ges_suite} combines the POSCAR-derived lattice schematic and DFT--Wannier-TB bands with integrated \(yyy\) shift current spectra from the three formulations and their local percentage differences relative to the Wilson loop result. The SAWF-TB and response quantities are recomputed from one symmetrized SAWF-gauge model, with the Hamiltonian, position, and derivative operators represented in the same gauge. The spectra use a \(200\times200\times1\) mesh and a 200-point \(0\)--\(4~\mathrm{eV}\) frequency grid. The Wilson loop QHC and shift vector maps use the subspaces \(\mathrm{V}=[19,20]\) and \(\mathrm{C}_1=[21,22]\) on the same centered \(400\times400\times1\) k-slice convention.

Panels (a) and (b) of Fig.~\ref{fig:ges_suite} show the lattice structure and the DFT and Wannier-TB dispersions on the same \(\Gamma\)--X--S--Y--\(\Gamma\) path with a common valence-band-maximum energy zero. This overlay provides an electronic structure comparison and is not used as a formal band-identity qualification.

For isolated bands, the projector trace expression reduces to the conventional shift current formula; when multiple bands are treated collectively, however, the subspace treatment built from Eq.~\eqref{eq:subspace_projector} is distinct from the single-band intermediates of Eq.~\eqref{eq:shift_sum_rule}. The three spectra in Fig.~\ref{fig:ges_suite}(c) have nearly indistinguishable line shapes under matched settings, so panel (d) uses the Wilson loop spectrum as the reference. The plotted percentage is the regularized function \(\delta_X^{yyy}\) defined in the caption. The absolute value in \(D_{\mathrm{W}}\) makes the denominator a reference-magnitude scale, and its \(5\%\) peak-response floor avoids a divergent or spuriously amplified percentage when \(\sigma_{\mathrm{W}}\) approaches or crosses zero. The maximum displayed magnitudes are \(2.65\%\) for the conventional comparison and \(0.0610\%\) for the projector comparison. For complementary global summaries, defining \(\epsilon_{L_2}(\mathrm{W},X)=\lVert\sigma_{\mathrm{W}}-\sigma_X\rVert_2/\lVert\sigma_{\mathrm{W}}\rVert_2\) and \(\epsilon_{\max}(\mathrm{W},X)=\max|\sigma_{\mathrm{W}}-\sigma_X|/\max|\sigma_{\mathrm{W}}|\), the projector comparison gives \(\epsilon_{L_2}=1.57\times10^{-4}\) and \(\epsilon_{\max}=2.90\times10^{-4}\), whereas the conventional comparison gives \(2.24\times10^{-3}\) and \(4.31\times10^{-3}\), respectively. Both measures therefore show that, with the Wilson loop spectrum as the common reference, the projector result is closer than the conventional result for this frozen dataset.

The Wilson loop formulation also provides the subspace-resolved k-space quantities in Fig.~\ref{fig:ges_suite}(e) and (f). Panel (e) resolves the imaginary QHC component \(C^{yyy}_{\mathrm{C}_1,\mathrm{V}}(\mathbf{k})\), while panel (f) shows the real shift vector \(R^{y;y}_{\mathrm{C}_1,\mathrm{V}}(\mathbf{k})\) for the same near-edge subspaces. Together, these maps resolve the transition geometry and its associated real-space displacement over the centered k slice. Their different dimensions, normalizations, and color scales preclude a pointwise comparison of magnitude or texture.

\begin{figure*}[!tbp]
\centering
\includegraphics[width=0.86\textwidth]{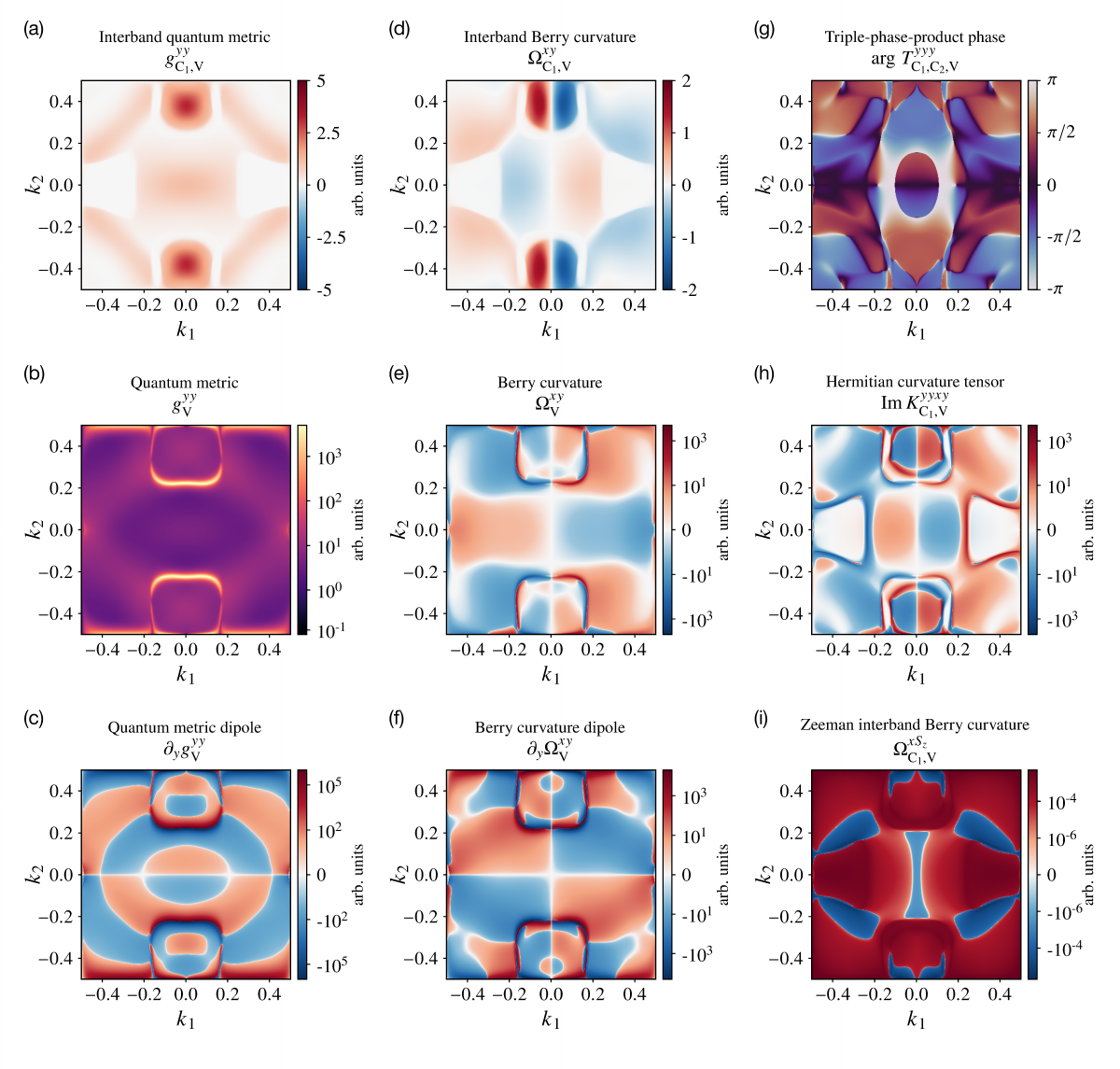}
\caption{GeS k-slice maps recomputed from one symmetrized SAWF-gauge model using direct Fourier interpolation on a \(400\times400\times1\) mesh, with \(E_F=-2.5~\mathrm{eV}\), Wannier-center Convention II, and \(\mathbf q=0\). The same model bundle supplies the Hamiltonian, position, spin, and derivative operators in a common gauge. The centered fractional coordinates satisfy \(\mathbf{k}=k_1\mathbf b_1+k_2\mathbf b_2\), with \(k_1,k_2\in[-0.5,0.5]\), and the subspaces are \(\mathrm{V}=[19,20]\), \(\mathrm{C}_1=[21,22]\), and \(\mathrm{C}_2=[23,24]\). The left column contains (a) \(g_{\mathrm{C}_1,\mathrm{V}}^{yy}\), (b) \(g_{\mathrm{V}}^{yy}\), and (c) \(\partial_{k_y}g_{\mathrm{V}}^{yy}\). The middle column contains (d) \(\Omega_{\mathrm{C}_1,\mathrm{V}}^{xy}\), (e) \(\Omega_{\mathrm{V}}^{xy}\), and (f) \(\partial_{k_y}\Omega_{\mathrm{V}}^{xy}\). The right column contains (g) \(\arg T_{\mathrm{C}_1,\mathrm{C}_2,\mathrm{V}}^{yyy}\), (h) \(\operatorname{Im}K_{\mathrm{C}_1,\mathrm{V}}^{yyxy}\), and (i) \(\Omega_{\mathrm{C}_1,\mathrm{V}}^{xS_z}\). Panels (a) and (d) use the zero-centered linear ranges \([-5,5]\) and \([-2,2]\), respectively. Panel (b) uses a linear-to-logarithmic color scale over \([0,5\times10^3]\); panels (c), (e), (f), (h), and (i) use signed logarithmic ranges of \(10^6\), \(5\times10^3\), \(2\times10^4\), \(5\times10^3\), and \(5\times10^{-3}\) in magnitude, respectively; and panel (g) uses the phase range \([-\pi,\pi]\). Each panel has its own color normalization and range, so colors do not provide a cross-panel comparison of magnitude. The displayed data have not been post-symmetrized.}
\label{fig:ges_qg_kslice_maps}
\end{figure*}

Different nonlinear optical responses probe pairwise, multistate, and spin-dependent aspects of quantum geometry, motivating their unified treatment within a common computational framework.
The quantum metric and Berry curvature resolve complementary amplitude and oriented-phase sectors of the same inter-subspace position-matrix geometry. Using the Hamiltonian-gauge position matrix element \(r^a_{nm}\) defined in Eq.~\eqref{eq:hamiltonian_position_decomposition}, the interband quantum geometric tensor between two generic band subspaces is \(Q_{A,B}^{ab}=N_A^{-1}N_B^{-1}\sum_{I\in A,J\in B}r^a_{IJ}r^b_{JI}\). Its real symmetric and imaginary antisymmetric parts give \(g_{A,B}^{ab}=\mathrm{Re}(Q_{A,B}^{ab}+Q_{A,B}^{ba})/2\) and \(\Omega_{A,B}^{ab}=-\mathrm{Im}(Q_{A,B}^{ab}-Q_{A,B}^{ba})\), respectively. For a generic subspace \(S\), \(g_S^{ab}=\sum_{I\in S,J\notin S}\mathrm{Re}(r^a_{IJ}r^b_{JI})\), and the corresponding derivative quantities are written as \(\partial_{k_c}g_S^{ab}\) and \(\partial_{k_c}\Omega_S^{ab}\). The metric maps in Fig.~\ref{fig:ges_qg_kslice_maps}(a) and (b) identify momentum-space regions of strong state mixing, whereas the Berry curvature maps in Fig.~\ref{fig:ges_qg_kslice_maps}(d) and (e) retain the oriented phase structure of the same interband geometry. Their derivatives in Fig.~\ref{fig:ges_qg_kslice_maps}(c) and (f) emphasize where these geometric weights vary most rapidly and hence where dipole-type nonlinear responses can acquire large local contributions after the response-specific occupation and energy weights are included. For the orthorhombic axes used here, \(k_1\parallel x\) and \(k_2\parallel y\). On the \(k_z=0\) plane, the \(C_{2v}\) operations \(m_x\) and \(C_{2y}\) both map \((k_1,k_2)\) to \((-k_1,k_2)\), while \(m_z\) leaves every in-plane k point fixed; time reversal maps \((k_1,k_2)\) to \((-k_1,-k_2)\). The \(yy\) metric components are even under these spatial operations and under time reversal, so the textures in panels (a) and (b) exhibit even parity in both \(k_1\) and \(k_2\). Their \(k_y\) derivative in panel (c) remains even in \(k_1\) but is odd in \(k_2\). The \(xy\) Berry curvature components change sign under \(m_x\) and \(C_{2y}\) and are odd under time reversal, giving textures that are odd in \(k_1\) and even in \(k_2\) in panels (d) and (e); the additional \(k_y\) derivative makes panel (f) odd in both axes. The subspace metric and metric-dipole maps in Fig.~\ref{fig:ges_qg_kslice_maps}(b) and (c) contain the internal contribution within the finite Wannier window. By contrast, the Berry curvature and Berry curvature dipole maps in Fig.~\ref{fig:ges_qg_kslice_maps}(e) and (f) include the Wannier-basis external-curvature contribution.

Beyond these pairwise tensors, the triple phase product and Hermitian curvature tensor probe multistate phase structure and the curvature of optical transition space, respectively. The triple phase product is the closed three-subspace product \(T_{A,B,C}^{abc}=(N_A N_B N_C)^{-1}\allowbreak\sum_{I\in A,J\in B,L\in C}r^a_{JI}r^b_{LJ}r^c_{IL}\). Its closed sequence of inter-subspace matrix elements retains relative phase information beyond pairwise metric and curvature tensors and enters gauge-invariant semiconductor Bloch equations and descriptions of high-harmonic optical response \cite{Liebscher2021Tellurium,Parks2023GISBE}, as well as bicircular-light-induced third-order multistate geometric currents \cite{Guo2025Bicircular}. The Hermitian curvature tensor treats optical transition dipole matrix elements as tangent vectors and measures the noncommutativity of their covariant transport. It is therefore the optical transition counterpart of Berry curvature and enters the third-order photovoltaic Hall response \cite{Ahn2022Riemannian}. For two disjoint subspaces \(A\) and \(B\), the subspace form implemented here is \(K_{AB}^{badc}=-\ii\Tr_{B}\!\left[r_{BA}^{b}\left(F_{A}^{dc}r_{AB}^{a}-r_{AB}^{a}F_{B}^{dc}\right)\right]\), where \(r_{AB}^{a}\) maps \(B\) to \(A\), \(r_{S}^{a}\) denotes the block of the already-defined Hamiltonian-gauge position matrix within subspace \(S\), and \(F_{S}^{dc}=\partial_{k_d}r_{S}^{c}-\partial_{k_c}r_{S}^{d}-\ii[r_{S}^{d},r_{S}^{c}]\) is its non-Abelian curvature. This expression uses the ordinary trace over \(B\), without division by \(N_A N_B\), and fixes the index order as \((b,a,d,c)\). Fig.~\ref{fig:ges_qg_kslice_maps}(h) shows \(\operatorname{Im}K_{\mathrm{C}_1\mathrm{V}}^{yyxy}\) and includes the Wannier-basis external-curvature contribution. All three Cartesian indices of \(T_{\mathrm{C}_1,\mathrm{C}_2,\mathrm{V}}^{yyy}\) are unchanged by \(m_x\) and \(C_{2y}\), making its phase even in \(k_1\). Time reversal gives \(T(-\kvec)=T(\kvec)^*\), so the texture in panel (g) exhibits odd \(k_2\) parity, with the relation understood modulo \(2\pi\) across the phase branch cut. By contrast, the single \(x\) index in \(K_{\mathrm{C}_1,\mathrm{V}}^{yyxy}\) makes its imaginary part odd in \(k_1\). The time-reversal complex-conjugation relation also makes this imaginary part odd under \(\kvec\rightarrow-\kvec\); together, the two constraints predict even \(k_2\) parity, as exhibited in panel (h).

Zeeman quantum geometry replaces one momentum-translation leg of the conventional quantum geometric tensor with a spin-rotation leg, yielding the k-resolved Zeeman Berry curvature component \(\Omega_{\mathrm{C}_1,\mathrm{V}}^{xS_z}\) shown in Fig.~\ref{fig:ges_qg_kslice_maps}(i). With \(s_{IJ}^{b}=\langle u_I|\hat{S}_{b}|u_J\rangle\), the subspace-averaged mixed position--spin tensor is \(\mathcal{Z}_{A,B}^{aS_b}=(N_A N_B)^{-1}\allowbreak\sum_{I\in A,\,J\in B}r_{JI}^{a}s_{IJ}^{b}=g_{A,B}^{aS_b}-\ii\Omega_{A,B}^{aS_b}/2\). Its anti-Hermitian part defines the Zeeman Berry curvature, \(\Omega_{A,B}^{aS_b}=-2\,\operatorname{Im}\mathcal{Z}_{A,B}^{aS_b}\) \cite{Xiang2025ZeemanQGT}. Under time reversal, the position matrix element is even while the spin matrix element is odd; antiunitarity and Kramers pairing therefore give \(\mathcal{Z}_{A,B}^{aS_b}(\kvec)=-[\mathcal{Z}_{\bar A,\bar B}^{aS_b}(-\kvec)]^*\). Consequently, the Zeeman quantum metric is time-reversal odd, whereas the Zeeman Berry curvature is time-reversal even. For Fig.~\ref{fig:ges_qg_kslice_maps}(i), \(A=\mathrm{C}_1\), \(B=\mathrm{V}\), \(a=x\), and \(b=z\); because both selected subspaces are closed under Kramers pairing, \(\Omega_{\mathrm{C}_1,\mathrm{V}}^{xS_z}(-\kvec)=\Omega_{\mathrm{C}_1,\mathrm{V}}^{xS_z}(\kvec)\). Under a spatial operation \(R\), \(r^x\) transforms as a polar-vector component, whereas \(S_z\) transforms as the axial-vector component \(\det(R)R_{zz}\). Their product is even under both \(m_x\) and \(C_{2y}\) and remains allowed under \(m_z\). The spatial relations therefore give even \(k_1\) parity, and the time-reversal-even character then gives even \(k_2\) parity. The calculated, non-post-symmetrized texture in panel (i) exhibits this predicted even--even pattern. The resulting map resolves the local coupling between orbital variation and spin texture in this spin--orbit-coupled band manifold and provides a geometric ingredient for the intrinsic gyrotropic magnetic current.

\FloatBarrier
\subsection{2H-\texorpdfstring{MoS$_2$}{MoS2} bilayer: photon drag response}

We use a 2H-MoS$_2$ bilayer to examine the finite-photon-momentum response of a centrosymmetric material. The bilayer has \(D_{3d}\) symmetry and therefore contains spatial inversion. For a spatially uniform optical field, this symmetry forbids the electric-dipole bulk photovoltaic response: the second-order current induced at \(q=0\) changes sign under inversion while the optical intensity does not. This makes the bilayer a useful benchmark for photon drag calculations, because any allowed response must be tied to the finite wave vector of light rather than to the ordinary \(q=0\) bulk photovoltaic mechanism.

Finite photon momentum changes this symmetry setting by introducing the polar vector \(\qvec\) into the optical transition. Photon drag injection current (PDIC) and photon drag shift current (PDSC) are therefore allowed channels even when the underlying crystal is inversion symmetric \cite{Shi2021PhotonDrag,Xie2025PhotonDragBPVE}. This material choice is also motivated by the recent PRL study of nonlinear circular photocurrent in 2D semiconductor MoS$_2$, where the observed helicity-dependent response was attributed to the circular photon drag effect rather than the circular photogalvanic effect \cite{Zhao2025MoS2CircularPhotocurrent}.

\begin{figure*}[t]
\centering
\includegraphics[width=\textwidth]{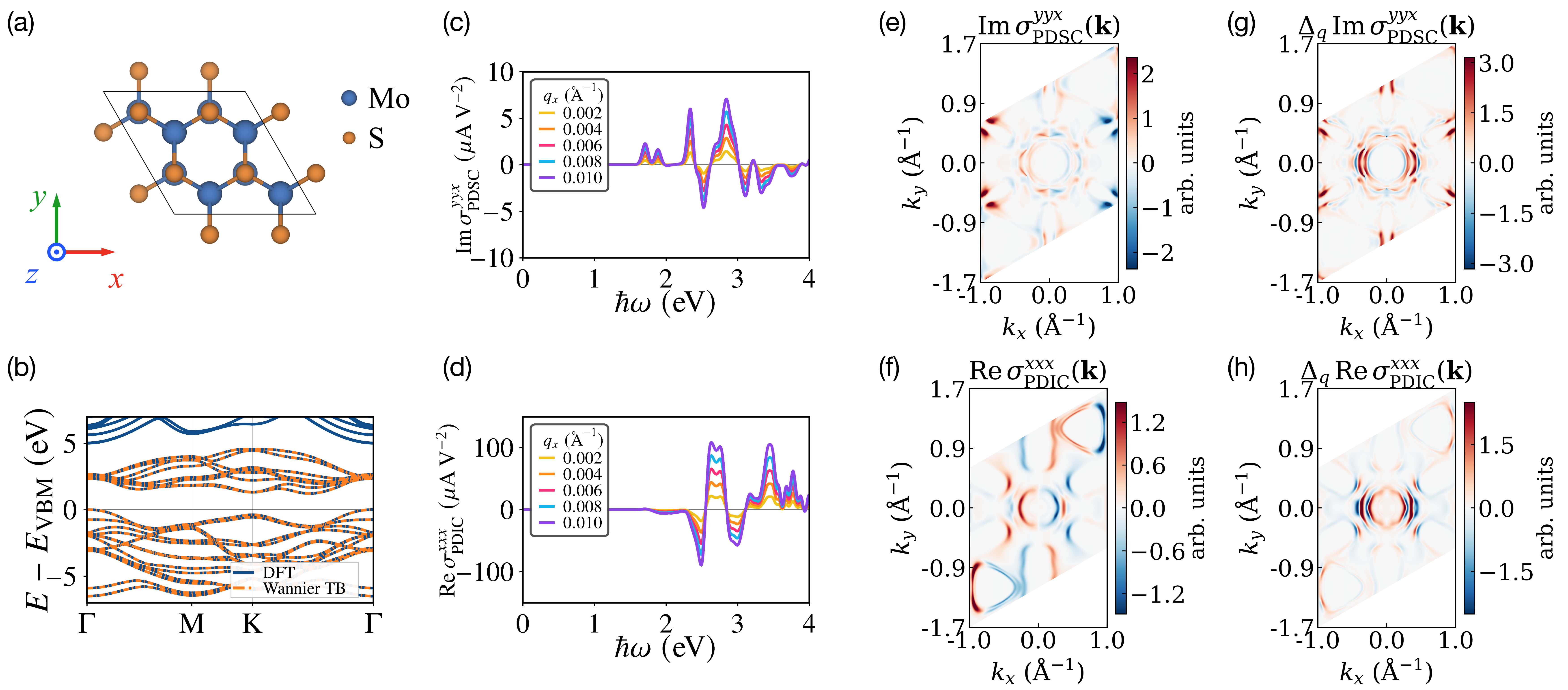}
\caption{2H-bilayer MoS$_2$ photon drag responses. (a) VESTA-rendered lattice structure and (b) DFT/SAWF band comparison. Panels (b)--(h) use the same SAWF model, with panels (c)--(h) calculated from its complete 11-operator bundle. (c) \(\mathrm{Im}\,\sigma^{yyx}_{\mathrm{PDSC}}\) and (d) \(\mathrm{Re}\,\sigma^{xxx}_{\mathrm{PDIC}}\) spectra for representative \(q_x\) values, both displayed as effective-bulk conductivities using \(L_z/d_{\mathrm{eff}}\) with \(d_{\mathrm{eff}}=0.950047554~\mathrm{nm}\); the PDIC spectra additionally use \(\tau=10^{-13}~\mathrm{s}\). (e) \(400\times400\times1\) reciprocal-space map of \(\mathrm{Im}\,\sigma^{yyx}_{\mathrm{PDSC}}(\mathbf{k})\), projected with the MoS$_2$ reciprocal lattice vectors, at \(q_x=0.01\,\mathrm{\mathring{A}}^{-1}\) and \(\hbar\omega=2.84~\mathrm{eV}\). (f) \(400\times400\times1\) reciprocal-space map of \(\mathrm{Re}\,\sigma^{xxx}_{\mathrm{PDIC}}(\mathbf{k})\), projected with the same reciprocal basis, at \(q_x=0.01\,\mathrm{\mathring{A}}^{-1}\) and \(\hbar\omega=2.75~\mathrm{eV}\). (g) and (h) Matched-parameter \(q_x\)-dependent maps obtained by subtracting the corresponding \(q_x=0\) response from panels (e) and (f), respectively, i.e. \(\Delta_q\sigma(\mathbf{k})=\sigma(q_x=0.01,\mathbf{k})-\sigma(q_x=0,\mathbf{k})\). Panels (e)--(h) show raw pointwise k-resolved K-slice kernels with an independent power-of-ten display scaling for each panel. They do not include Brillouin-zone quadrature weights, the effective-thickness rescaling, or the steady-state PDIC factor \(\tau\).}
\label{fig:mos2_suite}
\end{figure*}

Fig.~\ref{fig:mos2_suite} summarizes the MoS$_2$ photon drag calculations. Panels (c) and (d) show representative spectra for \(q_x=0.002,\ldots,0.010\,\mathrm{\mathring{A}}^{-1}\), making the growth of the finite-\(q_x\) response visible and selecting peak energies for the k-space maps. To match the effective-bulk convention of Fig.~\ref{fig:ges_suite}(c), the vacuum-diluted supercell responses are rescaled by \(L_z/d_{\mathrm{eff}}\), where \(L_z=3.06297918~\mathrm{nm}\) is the tight-binding supercell height and \(d_{\mathrm{eff}}=0.950047554~\mathrm{nm}\) is the outer S--S extent of the bilayer. In the present data, \(\mathrm{Im}\,\sigma^{yyx}_{\mathrm{PDSC}}\) reaches its representative \(q_x=0.010\,\mathrm{\mathring{A}}^{-1}\) peak at \(\hbar\omega=2.84~\mathrm{eV}\), with an effective-bulk value of about \(7.05~\mu\mathrm{A}/\mathrm{V}^2\). The corresponding \(\mathrm{Re}\,\sigma^{xxx}_{\mathrm{PDIC}}\) peak occurs at \(\hbar\omega=2.75~\mathrm{eV}\), with an effective-bulk value of about \(111.7~\mu\mathrm{A}/\mathrm{V}^2\).

The \(400\times400\times1\) k-space maps in Fig.~\ref{fig:mos2_suite}(e) and (f) show where these selected responses are concentrated in the Brillouin zone, using the Cartesian reciprocal plane defined by the MoS$_2$ reciprocal basis. The matched-parameter difference maps in panels (g) and (h) subtract the corresponding \(q_x=0\) maps from the \(q_x=0.010\,\mathrm{\mathring{A}}^{-1}\) maps. These difference panels isolate the photon-momentum-dependent part of the local response and separate it from structures that are already present in the \(q=0\) reference calculation.

The MoS$_2$ calculation also distinguishes the PDIC and PDSC implementation routes. PDIC is computed with conventional injection current vertices at finite \(\qvec\) and is controlled by the velocity imbalance of the optically connected states. PDSC uses the geometric loop formulation of shift current at finite \(\qvec\), where the shift vector information is encoded through the loop derivative. Both calculations use the same public task interface, metadata records, k-slice output format, and plotting workflow.

\subsection{Performance comparison with WannierBerri}

Dense Brillouin-zone integrations for nonlinear responses and k-resolved quantities demand high computational efficiency. The 2021 WannierBerri work identified the direct Fourier transform in the Wannier90 post-processing module \code{postw90.x} as a principal bottleneck and introduced a mixed scheme combining fast and slow Fourier transforms, which substantially accelerates Wannier interpolation. Together with its other optimizations, WannierBerri was reported to achieve speedups of several orders of magnitude over \code{postw90.x} \cite{Tsirkin2021WannierBerri}. WannierNLQG adopts the same mixed-Fourier strategy and further incorporates a series of optimizations across the numerical response workflow. To compare the computational efficiency of the two codes on a representative task supported by both, we selected the conventional shift current integral.

The time to solution was compared with WannierBerri 26.7.0, the latest version available at the time of testing, for three frozen material tasks. To prevent software-specific adaptive sampling or symmetry reduction from changing the benchmark workload, adaptive mesh refinement and symmetry-based reduction of the k mesh were disabled in WannierBerri, and WannierNLQG likewise evaluated the complete mesh. The reported speed ratios therefore do not include work reduction from the new response-symmetry path. Both programs used mixed FFT with identical \code{NKdiv} and \code{NKFFT} settings for each material. Here, \code{NKdiv} specifies the number of coarse k-space blocks along each reciprocal-space direction, whereas \code{NKFFT} specifies the FFT-grid size within each block; their componentwise product gives the full k mesh. For GeS and Nb\(_3\)I\(_8\), both programs used \code{NKdiv} \(=10\times10\times1\) and \code{NKFFT} \(=10\times10\times1\), yielding the \(100\times100\times1\) full k mesh. For GaAs, they used \code{NKdiv} \(=5\times5\times5\) and \code{NKFFT} \(=10\times10\times10\), yielding the \(50\times50\times50\) full k mesh. Both programs also used all bands and the same sampling in photon energy for the complete tensor workload of each material, at matched numbers of nominal compute units described below. Each program used a single public calculation call to write the complete \(3\times3\times3\) shift current tensor, comprising 27 Cartesian components. GeS, Nb\(_3\)I\(_8\), and GaAs used 200, 201, and 101 photon-energy samples, spanning \(0\)--\(4~\mathrm{eV}\), \(0\)--\(4~\mathrm{eV}\), and \(0\)--\(8~\mathrm{eV}\), respectively. The three material tasks therefore probe different workloads and are not treated as equal-FLOP benchmarks.

The resource count was \(N_{\mathrm{unit}}=4,8,12,16\), interpreted as MPI ranks for WannierNLQG and Ray workers for WannierBerri; Julia, OMP, MKL, OpenBLAS, VecLib, FFTW, and NumExpr thread counts were fixed to one. Each material--resource pair used one independent attempt containing eight fresh-process blocks in the order ABBA, BAAB, BAAB, ABBA, ABBA, BAAB, BAAB, ABBA, where A denotes WannierBerri and B denotes WannierNLQG. This yielded 16 warm task-time samples for each program at each resource count. The ratio for one block was the geometric mean of its two paired time ratios, and the reported speed ratio is the median over the eight block ratios. Confidence intervals were obtained from 10,000 bootstrap resamples stratified by ABBA/BAAB order. The warm interval excludes process launch, package import, and launcher overhead.

\begin{figure*}[!tbp]
\centering
\includegraphics[width=0.92\textwidth]{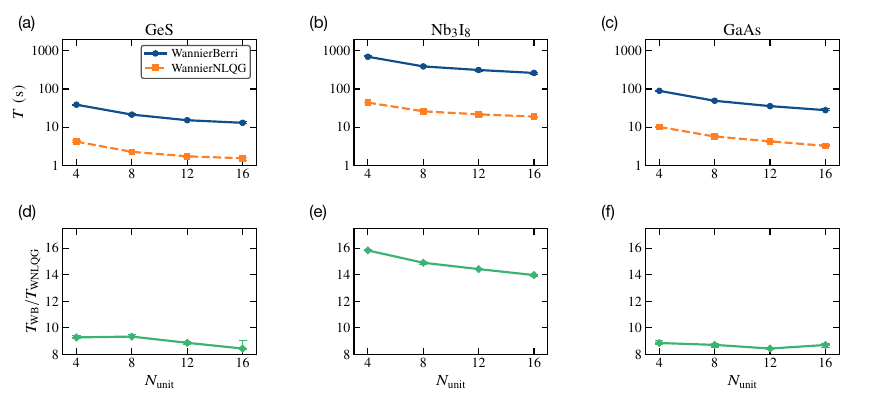}
\caption{Warm complete-task timing comparison for the full conventional shift current tensor of (a) and (d) GeS, (b) and (e) Nb\(_3\)I\(_8\), and (c) and (f) GaAs. Panels (a)--(c) show the median wall time from 16 samples, with min--max ranges, for WannierBerri 26.7.0 using Ray workers and the frozen WannierNLQG snapshot using MPI ranks. The time axes use logarithmic spacing and are labeled in seconds. Panels (d)--(f) show the median paired block ratio \(T_{\mathrm{WB}}/T_{\mathrm{WNLQG}}\), with order-stratified bootstrap 90\% confidence intervals. \(N_{\mathrm{unit}}\) denotes the number of Ray workers or MPI ranks. No samples or ratios are pooled across materials.}
\label{fig:performance_comparison}
\end{figure*}

Using the \(L_2\) norm introduced in the GeS example, the relative \(L_2\) difference between two aligned response vectors \(\mathbf{x}\) and \(\mathbf{y}\) is defined as \(L_{2,\mathrm{rel}}(\mathbf{x},\mathbf{y})=\lVert\mathbf{x}-\mathbf{y}\rVert_2/\max\!\left(\lVert\mathbf{x}\rVert_2,\lVert\mathbf{y}\rVert_2\right)\), where each vector contains the flattened response values over all sampled photon energies and tensor components. Fig.~\ref{fig:performance_comparison} shows a shorter WannierNLQG warm task time at every tested resource count. Across \(N_{\mathrm{unit}}=4\)--16, the observed paired speed ratios span \(8.44\)--\(9.34\times\) for GeS, \(13.97\)--\(15.83\times\) for Nb\(_3\)I\(_8\), and \(8.44\)--\(8.85\times\) for GaAs. All tested resource counts satisfied the validation checks for output completeness, finite values, frozen configuration, and within-program consistency relative to the four-unit result, using maximum-absolute and \(L_{2,\mathrm{rel}}\) tolerances of \(10^{-12}\) and \(10^{-10}\), respectively.
Cross-program numerical agreement was evaluated separately with a fixed canonical tensor mapping. The differences in the full tensor measured by \(L_{2,\mathrm{rel}}\) were \(0.563\%\) for GeS, \(0.080\%\) for Nb\(_3\)I\(_8\), and \(0.924\%\) for GaAs.

The observed difference in time to solution is consistent with several source-level design choices. WannierBerri constructs dense intermediates for a batch of k points before band-group and frequency contraction, whereas WannierNLQG processes individual k points with reusable, concretely typed workspaces and screens band pairs by occupations and resonant weights before the most expensive generalized-derivative, tensor, and frequency accumulations. This ordering avoids materializing large \(k\times n\times m\times27\) response arrays and reduces work on band pairs that cannot contribute. Persistent FFT buffers and plans, in-place kernels, static MPI block ownership, and deterministic reduction further reduce allocation and task-scheduling overhead.

Julia makes these algorithmic choices efficient through JIT specialization of concrete workspace types, compiled fine-grained loops, and direct in-place calls to BLAS and FFTW. The Fourier planning, matrix element workspaces, in-place execution, and MPI reduction are package-level infrastructure in WannierNLQG. Ordinary injection current and the shift current routes based on projectors, Wilson loops, and geometric loops consequently inherit the same optimization strategy.

\FloatBarrier

\section{Conclusions and outlook}
\label{sec:conclusions}

WannierNLQG provides a unified Julia framework for nonlinear optical responses and analysis of quantum geometry from Wannier tight-binding models. Its common \code{TaskConfig}, output, and metadata workflow supports ordinary and photon drag shift and injection currents in both integral and k-resolved calculations, enabling numerical settings and local response features to be cross-checked consistently. The conventional, projector, and generalized Wilson loop formulations, together with the geometric loop formulation at finite \(\qvec\), provide complementary routes to shift current quantities; the latter three use gauge-covariant subspace treatments for degenerate bands. The GeS and 2H-bilayer MoS$_2$ examples connect method comparisons, quantities in quantum geometry resolved by subspace, and response spectra at finite \(\qvec\) to their k-space origins. In frozen tests of the full conventional shift current tensor across three materials at matched numbers of nominal compute units, WannierNLQG achieved paired speedups in warm time to solution of \(8.44\)--\(15.83\times\) over WannierBerri 26.7.0; these tests used complete k meshes. Additional experimental workflows support symmetry-adapted Wannier construction, post-hoc symmetrization of the Hamiltonian and real-space operators, and irreducible-k-mesh integration in an invariant tensor basis; these expert paths remain in extended testing as material coverage grows.

Future development will extend the shared geometric infrastructure to additional nonlinear optical and transport responses, including thermoelectric, magnetoelectric, and optoelectric effects. Incorporating electron--phonon coupling and excitonic effects will broaden the physical processes represented in these calculations and support closer quantitative comparisons with experiment.

\section*{Code availability}

The source code of WannierNLQG corresponding to this article is available in the \href{https://github.com/ZhuochengLu/WannierNLQG}{GitHub repository} under the GNU General Public License version 2 only (GPL-2.0-only). Version 1.0.0 is preserved as a versioned \href{https://github.com/ZhuochengLu/WannierNLQG/releases/tag/v1.0.0}{GitHub release}; it contains the Julia source, documentation, and public examples.

\section*{Data availability}

The material-specific input files, representative numerical outputs, numerical metadata, and post-processing materials supporting the GeS and 2H-bilayer MoS$_2$ results are publicly available in Mendeley Data, Version 1 \cite{LuWannierNLQGData}. The public WannierNLQG v1.0.0 release provides the software documentation and synthetic examples. No human-subject, clinical, or otherwise restricted data were used in this study.

\section*{CRediT authorship contribution statement}

Zhuocheng Lu: Conceptualization; Data curation; Formal analysis; Investigation; Methodology; Software; Validation; Visualization; Writing -- original draft; Writing -- review \& editing. Zhichao Guo: Software, including development of the calculation code for the projector method; Formal analysis, including contributions to the generalization of the projector formulation to degenerate cases. Yuanyuan Xu: Formal analysis, including theoretical analysis of the Symmetrization module; Methodology; Software, including implementation of the module; Validation, including computational testing. Jiacheng Yao: Formal analysis, including theoretical analysis of the calculations of Berry curvature, quantum metric dipole, and quantum metric quadrupole. Hua Wang: Conceptualization; Funding acquisition; Project administration; Resources; Supervision; Writing -- review \& editing.

\section*{Declaration of competing interest}

The authors declare that they have no known competing financial interests or personal relationships that could have appeared to influence the work reported in this paper.

\section*{Acknowledgments}

H.W. acknowledges the support from the NSFC under Grants Nos. 12522411, 12474240, and 12304049, as well as the support from Fundamental Research Funds for the Central Universities.

\section*{AI and tool usage disclosure}

During the preparation of this work, the authors used OpenAI Codex with the GPT-5.6 model for language polishing and for assistance with performance optimization and refactoring of the software code. The prototype code and the scientific methodology were developed by the authors. All AI-assisted revisions were reviewed by the authors, who take full responsibility for the manuscript, the software, and the reported results.

\bibliographystyle{elsarticle-num}
\bibliography{references}

@article{KingSmith1993Polarization,
  author  = {King-Smith, R. D. and Vanderbilt, David},
  title   = {Theory of polarization of crystalline solids},
  journal = {Physical Review B},
  volume  = {47},
  number  = {3},
  pages   = {1651--1654},
  year    = {1993},
  doi     = {10.1103/PhysRevB.47.1651}
}

@article{Resta1994Polarization,
  author  = {Resta, Raffaele},
  title   = {Macroscopic polarization in crystalline dielectrics: the geometric phase approach},
  journal = {Reviews of Modern Physics},
  volume  = {66},
  number  = {3},
  pages   = {899--915},
  year    = {1994},
  doi     = {10.1103/RevModPhys.66.899}
}

@article{Marzari1997MLWF,
  author  = {Marzari, Nicola and Vanderbilt, David},
  title   = {Maximally localized generalized Wannier functions for composite energy bands},
  journal = {Physical Review B},
  volume  = {56},
  number  = {20},
  pages   = {12847--12865},
  year    = {1997},
  doi     = {10.1103/PhysRevB.56.12847}
}

@article{Souza2001EntangledWannier,
  author  = {Souza, Ivo and Marzari, Nicola and Vanderbilt, David},
  title   = {Maximally localized Wannier functions for entangled energy bands},
  journal = {Physical Review B},
  volume  = {65},
  number  = {3},
  pages   = {035109},
  year    = {2001},
  doi     = {10.1103/PhysRevB.65.035109}
}

@article{Sakuma2013SAWF,
  author  = {Sakuma, Rei},
  title   = {Symmetry-adapted Wannier functions in the maximal localization procedure},
  journal = {Physical Review B},
  volume  = {87},
  number  = {23},
  pages   = {235109},
  year    = {2013},
  doi     = {10.1103/PhysRevB.87.235109}
}

@article{Togo2024Spglib,
  author  = {Togo, Atsushi and Shinohara, Kohei and Tanaka, Isao},
  title   = {Spglib: a software library for crystal symmetry search},
  journal = {Science and Technology of Advanced Materials: Methods},
  volume  = {4},
  number  = {1},
  pages   = {2384822},
  year    = {2024},
  doi     = {10.1080/27660400.2024.2384822}
}

@article{Shinohara2023MagneticSymmetry,
  author  = {Shinohara, Kohei and Togo, Atsushi and Tanaka, Isao},
  title   = {Algorithms for magnetic symmetry operation search and identification of magnetic space group from magnetic crystal structure},
  journal = {Acta Crystallographica Section A},
  volume  = {79},
  number  = {5},
  pages   = {390--398},
  year    = {2023},
  doi     = {10.1107/S2053273323005016}
}

@article{Wang2006AHCWannier,
  author  = {Wang, Xinjie and Yates, Jonathan R. and Souza, Ivo and Vanderbilt, David},
  title   = {Ab initio calculation of the anomalous Hall conductivity by Wannier interpolation},
  journal = {Physical Review B},
  volume  = {74},
  number  = {19},
  pages   = {195118},
  year    = {2006},
  doi     = {10.1103/PhysRevB.74.195118}
}

@article{Yates2007WannierInterpolation,
  author  = {Yates, Jonathan R. and Wang, Xinjie and Vanderbilt, David and Souza, Ivo},
  title   = {Spectral and Fermi surface properties from Wannier interpolation},
  journal = {Physical Review B},
  volume  = {75},
  number  = {19},
  pages   = {195121},
  year    = {2007},
  doi     = {10.1103/PhysRevB.75.195121}
}

@article{Mostofi2008Wannier90,
  author  = {Mostofi, Arash A. and Yates, Jonathan R. and Lee, Young-Su and Souza, Ivo and Vanderbilt, David and Marzari, Nicola},
  title   = {wannier90: A tool for obtaining maximally-localised Wannier functions},
  journal = {Computer Physics Communications},
  volume  = {178},
  number  = {9},
  pages   = {685--699},
  year    = {2008},
  doi     = {10.1016/j.cpc.2007.11.016}
}

@article{Mostofi2014Wannier90Update,
  author  = {Mostofi, Arash A. and Yates, Jonathan R. and Pizzi, Giovanni and Lee, Young-Su and Souza, Ivo and Vanderbilt, David and Marzari, Nicola},
  title   = {An updated version of wannier90: A tool for obtaining maximally-localised Wannier functions},
  journal = {Computer Physics Communications},
  volume  = {185},
  number  = {8},
  pages   = {2309--2310},
  year    = {2014},
  doi     = {10.1016/j.cpc.2014.05.003}
}

@article{Pizzi2020Wannier90,
  author  = {Pizzi, Giovanni and Vitale, Valerio and Arita, Ryotaro and Blugel, Stefan and Freimuth, Frank and Geranton, Guillaume and Gibertini, Marco and Gresch, Dominik and Johnson, Charles and Koretsune, Takashi and Ibanez-Azpiroz, Julen and Lee, Hyungjun and Lihm, Jae-Mo and Marchand, Daniel and Marrazzo, Antimo and Mokrousov, Yuriy and Mustafa, Jamal I. and Nohara, Yoshiro and Nomura, Yusuke and Paulatto, Lorenzo and Ponce, Samuel and Ponweiser, Thomas and Qiao, Junfeng and Thole, Florian and Tsirkin, Stepan S. and Wierzbowska, Malgorzata and Marzari, Nicola and Vanderbilt, David and Souza, Ivo and Mostofi, Arash A. and Yates, Jonathan R.},
  title   = {Wannier90 as a community code: new features and applications},
  journal = {Journal of Physics: Condensed Matter},
  volume  = {32},
  number  = {16},
  pages   = {165902},
  year    = {2020},
  doi     = {10.1088/1361-648X/ab51ff}
}

@article{Sipe2000SecondOrder,
  author  = {Sipe, J. E. and Shkrebtii, A. I.},
  title   = {Second-order optical response in semiconductors},
  journal = {Physical Review B},
  volume  = {61},
  number  = {8},
  pages   = {5337--5352},
  year    = {2000},
  doi     = {10.1103/PhysRevB.61.5337}
}

@article{Wang2017NLOWannier,
  author  = {Wang, Chong and Liu, Xiaoyu and Kang, Lei and Gu, Bing-Lin and Xu, Yong and Duan, Wenhui},
  title   = {First-principles calculation of nonlinear optical responses by Wannier interpolation},
  journal = {Physical Review B},
  volume  = {96},
  number  = {11},
  pages   = {115147},
  year    = {2017},
  doi     = {10.1103/PhysRevB.96.115147}
}

@article{IbanezAzpiroz2018ShiftWannier,
  author  = {Ibanez-Azpiroz, Julen and Tsirkin, Stepan S. and Souza, Ivo},
  title   = {Ab initio calculation of the shift photocurrent by Wannier interpolation},
  journal = {Physical Review B},
  volume  = {97},
  number  = {24},
  pages   = {245143},
  year    = {2018},
  doi     = {10.1103/PhysRevB.97.245143}
}

@article{Rangel2017GeS,
  author  = {Rangel, Tonatiuh and Fregoso, Benjamin M. and Mendoza, Bernardo S. and Morimoto, Takahiro and Moore, Joel E. and Neaton, Jeffrey B.},
  title   = {Large Bulk Photovoltaic Effect and Spontaneous Polarization of Single-Layer Monochalcogenides},
  journal = {Physical Review Letters},
  volume  = {119},
  number  = {6},
  pages   = {067402},
  year    = {2017},
  doi     = {10.1103/PhysRevLett.119.067402}
}

@article{Wang2022WilsonLoop,
  author  = {Wang, Hua and Tang, Xiuyu and Xu, Haowei and Li, Ju and Qian, Xiaofeng},
  title   = {Generalized Wilson loop method for nonlinear light-matter interaction},
  journal = {npj Quantum Materials},
  volume  = {7},
  number  = {1},
  pages   = {61},
  year    = {2022},
  doi     = {10.1038/s41535-022-00472-4}
}

@article{Wang2017GiantSHG,
  author  = {Wang, Hua and Qian, Xiaofeng},
  title   = {Giant Optical Second Harmonic Generation in Two-Dimensional Multiferroics},
  journal = {Nano Letters},
  volume  = {17},
  number  = {8},
  pages   = {5027--5034},
  year    = {2017},
  doi     = {10.1021/acs.nanolett.7b02268}
}

@article{Wang2019FerroicityPhotocurrent,
  author  = {Wang, Hua and Qian, Xiaofeng},
  title   = {Ferroicity-driven nonlinear photocurrent switching in time-reversal invariant ferroic materials},
  journal = {Science Advances},
  volume  = {5},
  number  = {8},
  pages   = {eaav9743},
  year    = {2019},
  doi     = {10.1126/sciadv.aav9743}
}

@article{Wang2019FerroelectricNAHE,
  author  = {Wang, Hua and Qian, Xiaofeng},
  title   = {Ferroelectric nonlinear anomalous Hall effect in few-layer {WTe2}},
  journal = {npj Computational Materials},
  volume  = {5},
  number  = {1},
  pages   = {119},
  year    = {2019},
  doi     = {10.1038/s41524-019-0257-1}
}

@article{Wang2020SwitchablePhotocurrent,
  author  = {Wang, Hua and Qian, Xiaofeng},
  title   = {Electrically and magnetically switchable nonlinear photocurrent in {$\mathcal{PT}$}-symmetric magnetic topological quantum materials},
  journal = {npj Computational Materials},
  volume  = {6},
  number  = {1},
  pages   = {199},
  year    = {2020},
  doi     = {10.1038/s41524-020-00462-9}
}

@article{Xu2021BulkSpinPhotovoltaic,
  author  = {Xu, Haowei and Wang, Hua and Zhou, Jian and Li, Ju},
  title   = {Pure spin photocurrent in non-centrosymmetric crystals: bulk spin photovoltaic effect},
  journal = {Nature Communications},
  volume  = {12},
  number  = {1},
  pages   = {4330},
  year    = {2021},
  doi     = {10.1038/s41467-021-24541-7}
}

@article{Hu2025PhononQuantumGeometry,
  author  = {Hu, Jiaming and Li, Wenbin and Guo, Zhichao and Wang, Hua and Chang, Kai},
  title   = {Quantum Geometry in Phonon-Mediated Optical Responses},
  journal = {Physical Review Letters},
  volume  = {135},
  number  = {25},
  pages   = {256404},
  year    = {2025},
  doi     = {10.1103/y66p-kjj7}
}

@article{Wang2026GeodesicShiftVector,
  author  = {Wang, Hua and Chang, Kai},
  title   = {Geodesic Nature and Quantization of Shift Vector},
  journal = {Chinese Physics Letters},
  volume  = {43},
  number  = {2},
  pages   = {020703},
  year    = {2026},
  doi     = {10.1088/0256-307X/43/2/020703}
}

@article{Shi2021PhotonDrag,
  author  = {Shi, Li-kun and Zhang, Dong and Chang, Kai and Song, Justin C. W.},
  title   = {Geometric Photon-Drag Effect and Nonlinear Shift Current in Centrosymmetric Crystals},
  journal = {Physical Review Letters},
  volume  = {126},
  number  = {19},
  pages   = {197402},
  year    = {2021},
  doi     = {10.1103/PhysRevLett.126.197402}
}

@article{Xie2025PhotonDragBPVE,
  author  = {Xie, Ying-Ming and Nagaosa, Naoto},
  title   = {Photon-drag photovoltaic effects and quantum geometric nature},
  journal = {Proceedings of the National Academy of Sciences},
  volume  = {122},
  number  = {9},
  pages   = {e2424294122},
  year    = {2025},
  doi     = {10.1073/pnas.2424294122}
}

@article{Guo2025Projector,
  author  = {Guo, Zhichao and Lu, Zhuocheng and Wang, Hua},
  title   = {Projector method for nonlinear light-matter interactions and quantum geometry},
  journal = {Physical Review B},
  volume  = {112},
  number  = {23},
  pages   = {235115},
  year    = {2025},
  doi     = {10.1103/qd9q-hnfp}
}

@article{Avdoshkin2025MultistateGeometry,
  author  = {Avdoshkin, Alexander and Mitscherling, Johannes and Moore, Joel E.},
  title   = {Multistate Geometry of Shift Current and Polarization},
  journal = {Physical Review Letters},
  volume  = {135},
  number  = {6},
  pages   = {066901},
  year    = {2025},
  doi     = {10.1103/w761-8nf7}
}

@article{Ahn2022Riemannian,
  author  = {Ahn, Junyeong and Guo, Guang-Yu and Nagaosa, Naoto and Vishwanath, Ashvin},
  title   = {Riemannian geometry of resonant optical responses},
  journal = {Nature Physics},
  volume  = {18},
  number  = {3},
  pages   = {290--295},
  year    = {2022},
  doi     = {10.1038/s41567-021-01465-z}
}

@article{Xiang2025ZeemanQGT,
  author  = {Xiang, Longjun and Jia, Jinxiong and Xu, Fuming and Qiao, Zhenhua and Wang, Jian},
  title   = {Intrinsic Gyrotropic Magnetic Current from Zeeman Quantum Geometry},
  journal = {Physical Review Letters},
  volume  = {134},
  number  = {11},
  pages   = {116301},
  year    = {2025},
  doi     = {10.1103/PhysRevLett.134.116301}
}

@article{Liebscher2021Tellurium,
  author  = {Liebscher, Sven C. and Hagen, Maria K. and Hader, J{\"o}rg and Moloney, Jerome V. and Koch, Stephan W.},
  title   = {Microscopic theory for the incoherent resonant and coherent off-resonant optical response of tellurium},
  journal = {Physical Review B},
  volume  = {104},
  number  = {16},
  pages   = {165201},
  year    = {2021},
  doi     = {10.1103/PhysRevB.104.165201}
}

@article{Parks2023GISBE,
  author  = {Parks, A. M. and Moloney, J. V. and Brabec, T.},
  title   = {Gauge Invariant Formulation of the Semiconductor Bloch Equations},
  journal = {Physical Review Letters},
  volume  = {131},
  number  = {23},
  pages   = {236902},
  year    = {2023},
  doi     = {10.1103/PhysRevLett.131.236902}
}

@article{Guo2025Bicircular,
  author  = {Guo, Zhichao and Lu, Zhuocheng and Wang, Hua and Chang, Kai},
  title   = {Bicircular light-induced multistate geometric current},
  journal = {Physical Review B},
  volume  = {112},
  number  = {3},
  pages   = {035162},
  year    = {2025},
  doi     = {10.1103/7ytw-vyb7}
}

@misc{Lu2025GiantPhotonDrag,
  author        = {Lu, Zhuocheng and Qian, Zhuang and Guo, Zhichao and Shi, Likun and Liu, Shi and Wang, Hua and Chang, Kai},
  title         = {Giant Nonlinear Photon-Drag Currents in Moire Bilayers},
  year          = {2025},
  eprint        = {2511.16987},
  archivePrefix = {arXiv},
  note          = {arXiv:2511.16987v2, updated 2026-05-22}
}

@article{Gresch2017Z2Pack,
  author  = {Gresch, Dominik and Aut{\`e}s, Gabriel and Yazyev, Oleg V. and Troyer, Matthias and Vanderbilt, David and Bernevig, B. Andrei and Soluyanov, Alexey A.},
  title   = {{Z2Pack}: Numerical implementation of hybrid Wannier centers for identifying topological materials},
  journal = {Physical Review B},
  volume  = {95},
  number  = {7},
  pages   = {075146},
  year    = {2017},
  doi     = {10.1103/PhysRevB.95.075146}
}

@article{Lee2023EPW,
  author  = {Lee, Hyungjun and Ponc{\'e}, Samuel and Bushick, Kyle and Hajinazar, Samad and Lafuente-Bartolome, Jon and Leveillee, Joshua and Lian, Chao and Lihm, Jae-Mo and Macheda, Francesco and Mori, Hitoshi and Paudyal, Hari and Sio, Weng Hong and Tiwari, Sabyasachi and Zacharias, Marios and Zhang, Xiao and Bonini, Nicola and Kioupakis, Emmanouil and Margine, Elena R. and Giustino, Feliciano},
  title   = {Electron--phonon physics from first principles using the {EPW} code},
  journal = {npj Computational Materials},
  volume  = {9},
  number  = {1},
  pages   = {156},
  year    = {2023},
  doi     = {10.1038/s41524-023-01107-3}
}

@article{Vitale2020AutomatedWannierisation,
  author  = {Vitale, Valerio and Pizzi, Giovanni and Marrazzo, Antimo and Yates, Jonathan R. and Marzari, Nicola and Mostofi, Arash A.},
  title   = {Automated high-throughput Wannierisation},
  journal = {npj Computational Materials},
  volume  = {6},
  number  = {1},
  pages   = {66},
  year    = {2020},
  doi     = {10.1038/s41524-020-0312-y}
}

@article{Qiao2023ProjectabilityDisentanglement,
  author  = {Qiao, Junfeng and Pizzi, Giovanni and Marzari, Nicola},
  title   = {Projectability disentanglement for accurate and automated electronic-structure Hamiltonians},
  journal = {npj Computational Materials},
  volume  = {9},
  number  = {1},
  pages   = {208},
  year    = {2023},
  doi     = {10.1038/s41524-023-01146-w}
}

@article{Marrazzo2024WannierEcosystem,
  author  = {Marrazzo, Antimo and Beck, Sophie and Margine, Elena R. and Marzari, Nicola and Mostofi, Arash A. and Qiao, Junfeng and Souza, Ivo and Tsirkin, Stepan S. and Yates, Jonathan R. and Pizzi, Giovanni},
  title   = {Wannier-function software ecosystem for materials simulations},
  journal = {Reviews of Modern Physics},
  volume  = {96},
  number  = {4},
  pages   = {045008},
  year    = {2024},
  doi     = {10.1103/RevModPhys.96.045008}
}

@article{Tsirkin2021WannierBerri,
  author  = {Tsirkin, Stepan S.},
  title   = {High performance Wannier interpolation of Berry curvature and related quantities with WannierBerri code},
  journal = {npj Computational Materials},
  volume  = {7},
  number  = {1},
  pages   = {33},
  year    = {2021},
  doi     = {10.1038/s41524-021-00498-5}
}

@misc{WannierBerriDocumentation,
  title        = {{WannierBerri} documentation},
  howpublished = {\url{https://docs.wannier-berri.org/}},
  note         = {Accessed 4 September 2026},
  year         = {2026}
}

@article{Zhi2022WannSymm,
  author  = {Zhi, Guo-Xiang and Xu, Chenchao and Wu, Si-Qi and Ning, Fanlong and Cao, Chao},
  title   = {WannSymm: A symmetry analysis code for Wannier orbitals},
  journal = {Computer Physics Communications},
  volume  = {271},
  pages   = {108196},
  year    = {2022},
  doi     = {10.1016/j.cpc.2021.108196}
}

@article{Wu2018WannierTools,
  author  = {Wu, QuanSheng and Zhang, ShengNan and Song, Hai-Feng and Troyer, Matthias and Soluyanov, Alexey A.},
  title   = {WannierTools: An open-source software package for novel topological materials},
  journal = {Computer Physics Communications},
  volume  = {224},
  pages   = {405--416},
  year    = {2018},
  doi     = {10.1016/j.cpc.2017.09.033}
}

@article{Zhao2025MoS2CircularPhotocurrent,
  author  = {Zhao, Yanchong and Chen, Fengyu and Liang, Jing and Bahramy, Mohammad Saeed and Yang, Mingwei and Guang, Yao and Li, Xiaomei and Wei, Zheng and Tang, Jian and Zhao, Jiaojiao and Liao, Mengzhou and Shen, Cheng and Wang, Qinqin and Yang, Rong and Watanabe, Kenji and Taniguchi, Takashi and Huang, Zhiheng and Shi, Dongxia and Liu, Kaihui and Sun, Zhipei and Feng, Ji and Du, Luojun and Zhang, Guangyu},
  title   = {Origin of Nonlinear Circular Photocurrent in 2D Semiconductor MoS2},
  journal = {Physical Review Letters},
  volume  = {134},
  number  = {8},
  pages   = {086201},
  year    = {2025},
  doi     = {10.1103/PhysRevLett.134.086201}
}

@article{torma2022superconductivity,
  title={Superconductivity, superfluidity and quantum geometry in twisted multilayer systems},
  author={T{\"o}rm{\"a}, P{\"a}ivi and Peotta, Sebastiano and Bernevig, Bogdan A},
  journal={Nature Reviews Physics},
  volume={4},
  number={8},
  pages={528--542},
  year={2022},
  publisher={Nature Publishing Group UK London}
}

@misc{LuWannierNLQGData,
  author    = {Lu, Zhuocheng and Guo, Zhichao and Xu, Yuanyuan and Yao, Jiacheng and Wang, Hua},
  title     = {Reproducibility data for {WannierNLQG}: A {Julia} package for nonlinear optical responses and quantum geometry from {Wannier} tight-binding models},
  howpublished = {Mendeley Data},
  year      = {2026},
  note      = {Version 1},
  doi       = {10.17632/v2s2j336vh.1}
}

@article{liu2025quantum,
  title={Quantum geometry in condensed matter},
  author={Liu, Tianyu and Qiang, Xiao-Bin and Lu, Hai-Zhou and Xie, XC},
  journal={National Science Review},
  volume={12},
  number={3},
  pages={nwae334},
  year={2025},
  publisher={Oxford University Press}
}

\end{document}